\documentclass[preprintnumbers,amsmath,amssymb,superscriptaddress,twocolumn]{revtex4}

\usepackage{calc}
\usepackage{graphicx}
\usepackage{bm}
\usepackage{color}

\newcommand{\pa}{\partial}
\newcommand{\pd}[2]{\frac{\partial #1}{\partial #2}}
\newcommand{\pdds}[2]{\frac{\partial^2 #1}{\partial #2^2}}

\newcommand{\braket}[2]{\bigl\langle#1\bigl|\bigr.#2\bigr\rangle}

\newcommand{\nm}[1]{\left|#1\right|}

\newcommand{\tp}[1]{\tilde{\theta}}

\begin{document}

\title{Exact mean exit time for surface-mediated diffusion}

\author{J.-F. Rupprecht}
\affiliation{Laboratoire de Physique Th\'eorique de la Mati\`ere Condens\'ee
(UMR 7600), case courrier 121, Universit\'e Paris 6, 4 Place Jussieu, 75255
Paris Cedex}

\author{O. B\'enichou}
\affiliation{Laboratoire de Physique Th\'eorique de la Mati\`ere Condens\'ee
(UMR 7600), case courrier 121, Universit\'e Paris 6, 4 Place Jussieu, 75255
Paris Cedex}

\author{D. S. Grebenkov}
\affiliation{Laboratoire de Physique de la Mati\`ere Condens\'ee (UMR 7643),
CNRS -- Ecole Polytechnique, F-91128 Palaiseau Cedex France}

\author{R. Voituriez}
\affiliation{Laboratoire de Physique Th\'eorique de la Mati\`ere Condens\'ee
(UMR 7600), case courrier 121, Universit\'e Paris 6, 4 Place Jussieu, 75255
Paris Cedex}

\date{\today}

\begin{abstract}
We present an exact expression for the mean exit time through the cap of
a confining sphere for particles alternating phases of surface and of bulk diffusion.  The present approach is based on an integral equation which can be solved analytically.  
In contrast to the statement of Berezhkovskii and Barzykin [J. Chem. Phys. {\bf 136}, 54115 (2012)],  we show that the mean
exit time can be optimized with respect to the desorption rate, under
analytically determined criteria.
\end{abstract}

\maketitle

\section{Introduction}

Surface-mediated processes, in which a particle randomly alternates
between surface and bulk diffusions, are relevant for various chemical
and biochemical processes such as reactions in porous media
or trafficking in living cells
\cite{Walder:2011,Chechkin:2009,Schuss:2007,Alberts:2002, 
Bychuk:1995, bond, Astumian:1985,Berg:1981,Sano:1981, Adam:1968}.
Recently, it has been shown theoretically that intermittent dynamics
between two diffusive phases can lead to faster kinetics.  The
kinetics was characterized by the mean first passage time (MFPT) of
diffusing particles to a fixed reactant \cite{Benichou:2011b}.

As a representative example of confined interfacial kinetics, the case
of a particle alternating bulk and surface diffusion over a sphere
which contains a reactive cap, was considered in
\cite{Benichou:2011a,Benichou:2010,Calandre:2012}.  The reactive cap
can also be interpreted as a hole, in which case the MFPT is referred
to as the mean exit time  \cite{Schuss:2012,ward:2010,al:2011,Grigoriev:2002a}. 
While the desorption rate from the surface is
independent of any geometrical parameter, the switching dynamics from
the bulk to the surface is determined by the statistics of returns to
the sphere.  These statistics strongly depend on the boundary behavior
of the process.  After each desorption event, it was assumed in
\cite{Benichou:2011a,Benichou:2010} that the particle was ejected
into the bulk at a distance $a > 0$ that was required to avoid an
immediate re-adsorption on the perfectly adsorbing sphere.  A similar
description was used in discrete square lattice versions of this
model, in which the lattice spacing played the role of the cut-off
distance $a$ \cite{Rojo:2011,Rojo:2012}.  In these studies, the MFPT
was found, under certain conditions, to be an optimizable function of
the desorption rate.  The favorable effect of desorptions was
attributed in \cite{Benichou:2011a} to the fact that bulk excursions
reduce the time wasted due to the recurrence of surface Brownian
motion, by bringing particles through the bulk to unvisited regions of
the sphere.  Previous mean-field treatments which ignored spatial
correlations, missed this possible optimum \cite{Oshanin:2010}.  In
the case of a uniformly semi-reflecting sphere, {\it including the
target}, the MFPT was also found to be an optimizable function of the
desorption rate even for a distance of ejection $a$ set to zero
\cite{Rupprecht:2012a}.  

Recently, a coarse-grained approach to the surface-mediated search for
a perfectly adsorbing target in an otherwise semi-reflecting sphere
was considered in \cite{berezhkovskii:054115}.  This model is
relevant in numerous real situations in which the particle
reactivity with the target is not related to its affinity with the
rest of the surface, as it is obviously the case in particular for
exit problems. Relying on an elegant first order kinetics scheme, 
concise approximate expressions were proposed in \cite{berezhkovskii:054115} 
for the spatially averaged MFPTs (called here global MFPTs and denoted GMFPTs) 
for a uniform distribution of starting points over either the sphere surface or the bulk. 
These averages will be called hereafter surface and bulk GMFPTs.  
This approach, which treats the bulk and surface diffusive
phases as two effective states coupled by first order kinetic
equations, led to a monotonic GMFPT as a function of the desorption
rate.  This striking difference with Ref. \cite{Benichou:2011a} was attributed in \cite{berezhkovskii:054115} to the non-locality of
the desorption process, in which the instantaneous ejection at a non
zero distance $a$ implied a violation of the detailed balance
condition.

In this article we clarify this puzzling situation and address the
question of the optimality of the GMFPTs as a function of the
desorption rate for the mixed boundary condition of
\cite{berezhkovskii:054115}.  More precisely, (i) we provide an exact
solution for the MFPT; (ii) we prove that the surface GMFPT can still be
optimized with respect to the desorption rate, under analytically
determined criteria; (iii) we compare our results with the
coarse-grained approach of \cite{berezhkovskii:054115}, which is shown
to be accurate only in a limited region of the parameter space.

\begin{figure}[h!]
 \centering
 \includegraphics[width=5cm]{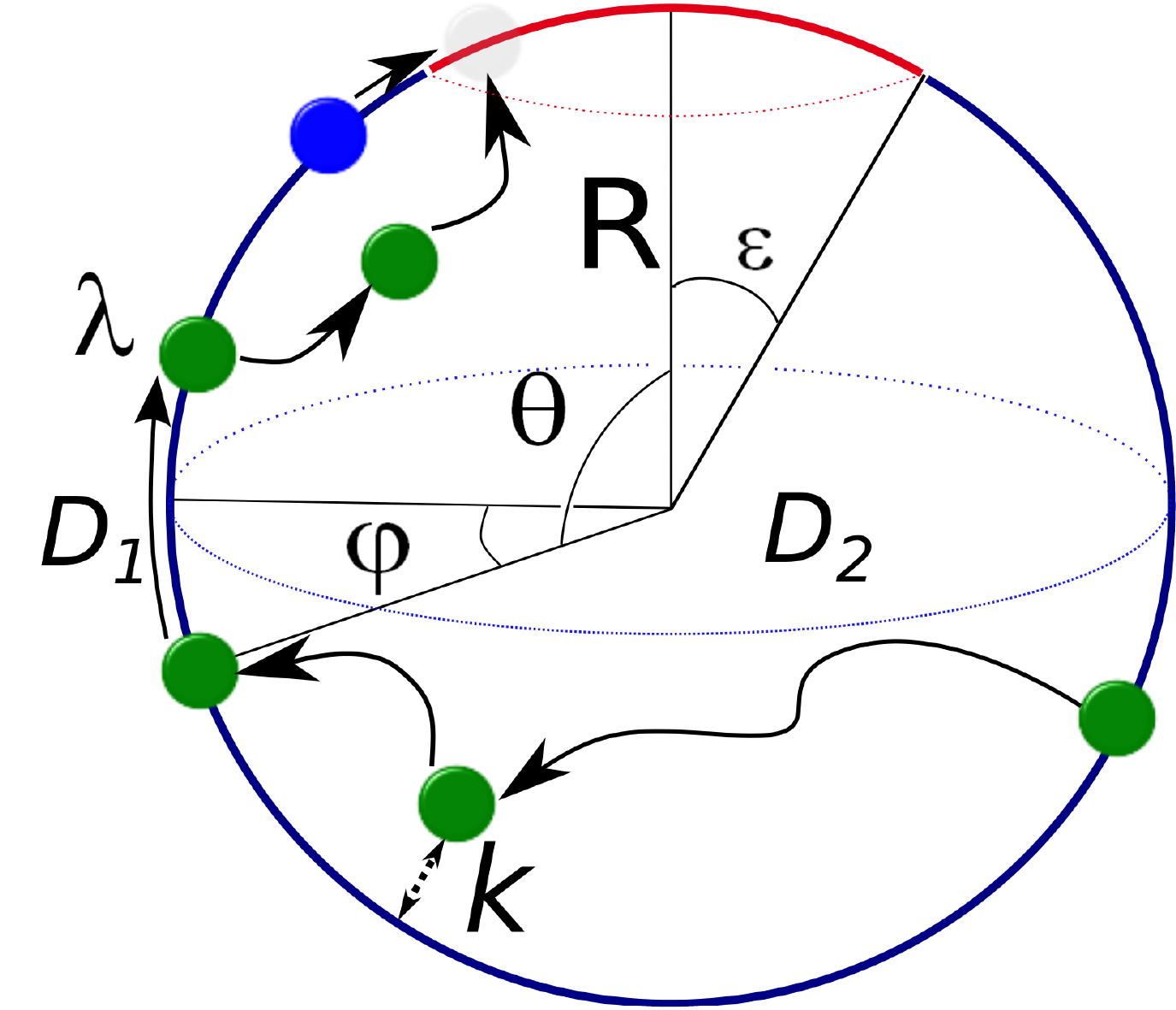}
\caption{
(Color online).  Model of surface-mediated diffusion in a sphere of
radius $R$ containing a perfectly adsorbing cap (a target).  For
finite values of $k$, particles (represented as small spheres)
diffusing in the bulk randomly bind to or bounce from the remaining
(non-adsorbing) part of the sphere surface (outside the target cap).
The target is reached either from the bulk (shown by the green
particle), or from the surface (shown by the blue particle).}
\label{fig:model}
\end{figure}

\section{The model} \label{sec:model}

We consider particles diffusing in a three-dimensional spherical
cavity of radius $R$ (see Fig. \ref{fig:model}) switching between
phases of surface diffusion with diffusion coefficient $D_1$ and
phases of bulk diffusion with diffusion coefficient $D_2$.  The time
spent on the surface is assumed to follow an exponential law with
desorption rate $\lambda$.

The target is a cap defined as the portion of the sphere
$\theta\in[0,\epsilon]$, where $\theta$ is the elevation angle in
spherical coordinates. The target is considered as perfectly
reactive, i.e. particles react as soon as they encounter the target
for the first time. We will focus here on the mean 
first passage time (MFPT) on the target. The surface GMFPT $\langle t_1 \rangle$ is defined here
as the average of the MFPT over a uniform distribution of the starting
points on the sphere surface, \textit{including the target}.  In turn,
in Ref. \cite{berezhkovskii:054115}, the MFPT was averaged over the
starting points outside the target; we denote this average as $\langle t_1
\rangle_{[1]}$.  The two averages are related by $\langle t_1 \rangle=A
\langle t_1 \rangle_{[1]}$, where $A$ is a geometrical factor which
is written as
\begin{equation}
\label{eq:A}
A = \frac{1}{1+\tan^2(\epsilon/2)}. 
\end{equation}

The non-reactive part of the sphere is considered as semi-reflecting:
a particle reaching by bulk diffusion the non-reactive part of the
sphere surface gets randomly either adsorbed or reflected back in the
bulk.  The probability for binding to the surface is an increasing
function of the adsorption coefficient $k$, which will be precisely
defined through the radiative boundary condition for the MFPT in
Eq. (\ref{eq:bcreflective}).  In particular, $k = \infty$ (resp. $k =
0$) corresponds to a perfectly adsorbing (resp. reflecting) boundary.
Notice that the model that we consider here is exactly the same as in
\cite{berezhkovskii:054115}.

The limits $\lambda=0$ and $\lambda=\infty$ correspond to simple
situations.  In the case $\lambda = 0$, particles are trapped on the
surface until they reach the target, and the exact expression for the surface
GMFPT is
\cite{Linderman:1951}:
\begin{equation} \label{eq:taus3D}
t_s = \langle t_1 \rangle_{\lambda = 0} = \frac{R^2}{D_1} \left(\ln\left(\frac{2}{1-\cos \epsilon}\right) - \frac{1+\cos \epsilon}{2}\right).
\end{equation}
The limit $\lambda = \infty$ is equivalent to a purely reflecting
boundary.  The asymptotic behavior for the narrow escape limit
$\epsilon \ll 1$ is given in \cite{Singer:2006b}:
\begin{equation} \label{eq:taub3D}
t_b = \langle t_1 \rangle_{\lambda =\infty} = \frac{\pi R^2}{3 D_2 \epsilon} \biggl( 1 + \epsilon \ln(\epsilon) + O(\epsilon) \biggr).
\end{equation}  
Notice that the bulk and surface GMFPTs diverge as $\epsilon$ tends to
zero, as it can be seen in particular in the above limits.  Indeed a
point-like target ($\epsilon = 0$) is detectable neither by bulk (3D)
excursions nor by surface (2D) diffusion.

\section{Exact solution} \label{sec:simplegeom}

\subsection{Basic equations} \label{eq:basiceq}

Using a standard formalism of backward equations \cite{Gardiner:2004}, we derive 
the diffusive equations (\ref{eq:temps1},\ref{eq:temps2}) 
and the appropriate boundary conditions (\ref{eq:bcadsorb},\ref{eq:bcreflective}) 
satisfied by the MFPT, which we proceed to solve in the next section.

We define $p((r,\theta,\phi),s|(\bm{x'},\phi'),t')$
(resp. $p((\theta,\phi),s|\bm{x'},t')$) as the probability for a
particle located at time $t'$ at $\bm{x'}$ -- with either $\bm{x'} =
r',\theta'$ into the bulk or $\bm{x'} = \theta'$ on the sphere -- to
reach at time $s > t'$ the point $(r,\theta,\phi)$ in the bulk
(resp. the point $(\theta,\phi)$ on the surface).  In what follows, we
omit the azimuthal coordinates by referring to probabilities averaged
over the initial and final azimuthal angles:
\begin{align} \nonumber
p((r,\theta),s|\bm{x'},t') &\equiv \int^{2 \pi}_{0} d\phi \int^{2 \pi}_{0} d\phi'  p((r,\theta,\phi),s|(\bm{x'},\phi'),t').
\end{align}
For convenience, in this section we will use the shorthand notations
\begin{align}
 p((r,\theta),s|\bm{x'},t') \equiv p(r,\theta), \qquad p(\theta,s|\bm{x'},t') \equiv p(\theta).
\end{align}
Conversely, we define the conditional probabilities:
$p(\bm{\bar{x}},\bar{t}|(r,\theta),s)
\equiv \bar{p}(r,\theta)$ and $p(\bm{\bar{x}},\bar{t}|\theta,s) \equiv
\bar{p}(\theta)$, with $\bar{t} > s$ and $\bar{x} = (\bar{r},\bar{\theta})$ in
the bulk or $\bar{x} = \bar{\theta}$ on the sphere. \\

In the following, the Laplace operator is $\Delta_{(r,\theta)} =
\Delta_{r} + \Delta_{\theta}/r^2$ where $\Delta_{r}$ and
$\Delta_{\theta}$ are
\begin{equation} \label{eq:diffusionoperators3D}
\Delta_{r} = \pdds{}{r} + \frac{2}{r} \pd{}{r} , \qquad 
\Delta_{\theta} = \frac{1}{\sin \theta} \; \pa_{\theta}\; ( \sin\theta \; \pa_{\theta}).
\end{equation}

\subsubsection{Diffusion equations}

For the process under study the conditional probability $p(r,\theta)$
satisfies the \textit{forward} diffusion equations, for all $s>t'$,
\cite{berezhkovskii:054115},
\begin{align}
\pd{ p(r,\theta)}{s} &= D_2 \ \Delta_{(r,\theta)} \ p(r,\theta)  \label{eq:forwarddiffb} \\
 \pd{ p(\theta)}{s} &= 
\frac{D_{1}}{R^{2}} \Delta_{\theta} p(\theta) - \lambda \ p(\theta) + k D_2 \ p(R,\theta)
\label{eq:forwarddiffsurface}.
\end{align}
Each term in the right-hand side of Eq. (\ref{eq:forwarddiffsurface}) has a
straightforward physical interpretation ; these are (from left to
right) (i) the diffusion within the surface state; (ii) desorption
events with a constant rate $\lambda$; (iii) adsorption events with a
success rate quantified by $k$ (\cite{Rupprecht:2012a}).

Equivalently to
Eqs. (\ref{eq:forwarddiffb}) and (\ref{eq:forwarddiffsurface}), the
conditional probabilities satisfy the backward equations, for all
$t>s'$,
\cite{Gardiner:2004}
\begin{align}
  \label{eq:Chapfrom2}
 \pd{ \bar{p}(r,\theta)}{s} &= - D_{2} \ \Delta_{(r,\theta)} \ \bar{p}(r,\theta). \\
  \label{eq:Chapfrom1}
  \pd{ \bar{p}(\theta)}{s} &= - \frac{D_{1}}{R^2} \Delta_{\theta} \ \bar{p}(\theta) + 
  \lambda  \biggl\{\bar{p}(\theta) - \bar{p}(R,\theta)\biggr\}
\end{align}
These backward diffusion equations are commonly used to determine 
first passage time observables \cite{Redner:2001a}. 

We define $t_1(\theta)$ as the MFPT for particles started on the
sphere at the angle $\theta$, and $t_2(r,\theta)$ stands for the MFPT
for particles started at the bulk point $(r,\theta)$ (the second
angular coordinate $\phi$ is irrelevant due to the symmetry and thus
ignored). The MFPTs are expressed in terms of the conditional
probabilities through the relations \cite{Gardiner:2004}
\begin{align} 
\label{def:t1}
 t_{1}(\theta)  &\equiv \int^{\infty}_{0} dt 
		\left( \int^{\pi}_{0} 2 \pi R^{2} \sin \tilde{\theta} d\tilde{\theta} \ p(\tilde{\theta},t|\theta,0) \right. \nonumber \\
		&+ \left. \int_{\mathcal{S}} 2 \pi r^2 dr \sin \tilde{\theta} d\tilde{\theta} \ p(r,\tilde{\theta},t|\theta,0)\right), \\
\label{def:t2}
 t_{2}(r,\theta)&\equiv \int^{\infty}_{0}  dt 
		   \left( \int^{\pi}_{0} 2 \pi R^{2} \sin \tilde{\theta} d\tilde{\theta} \ p(\tilde{\theta},t|r,\theta,0) \right. \nonumber \\
		&+ \left. \int_{\mathcal{S}} 2 \pi r^2 dr \sin \tilde{\theta} d\tilde{\theta} \ p(r,\tilde{\theta},t|r,\theta,0)\right),
\end{align}
where $\mathcal{S} \equiv (0,R)\times(0,2\pi)$.

Substituting the latter relations (\ref{def:t1}) and (\ref{def:t2}) into
Eqs. (\ref{eq:Chapfrom2},\ref{eq:Chapfrom1}), one can show that the
MFPTs satisfy the set of equations
\begin{align}  
\label{eq:temps1}
\frac{D_1}{R^2} \Delta_{\theta} t_1(\theta) + 
\lambda  \left(t_2(R,\theta) - t_1(\theta) \right) &= - 1  \quad (\epsilon < \theta < \pi), \\
\label{eq:temps2} 
D_2 \left( \Delta_{r} + \frac{\Delta_{\theta}}{r^2}  \right) \; t_2(r,\theta) &= -1 \quad
 ((r,\theta) \in \mathcal{S}).
\end{align}
In the next section we specify the appropriate boundary conditions for the MFPT.

\subsubsection{Boundary conditions} \label{sec:bc}

We justify that the B.C. for the process defined in
Sec. \ref{sec:model} are:

(i) the Dirichlet boundary condition
\begin{equation} 
\label{eq:bcadsorb} 
t_1(\theta) = 0    \qquad (0 \leq \theta \leq \epsilon),
\end{equation}
which expresses that the search process is stopped on the target. 

(ii) the mixed boundary condition
\begin{align} 
\label{eq:bcreflective}
t_2(R,\theta) = \begin{cases} 0  \hskip 33mm (0 \leq \theta \leq \epsilon), \cr
\displaystyle t_1(\theta) - \frac{1}{k} \pd{t_2}{r}_{\lvert {\bf{r}}=(R,\theta)} \hskip 4mm (\epsilon < \theta \leq  \pi), \end{cases}
\end{align}
in which the first relation expresses perfect adsorption on the target
($\theta \in [0,\epsilon]$) while the second relation (radiative B.C.)
implements an imperfect adsorption process on the rest of the surface
($\theta \in [\epsilon,\pi]$).  The mixed boundary condition (\ref{eq:bcreflective})
is the major difference of the present model from the previously studied case
\cite{Rupprecht:2012a}, in which the radiative B.C. was imposed over
the whole boundary, including the target.  Although this modification
may seem minor, the difference on the target reactivity has a drastic
effect on the surface GMFPT for short adsorption times ($\lambda \gg
D_1/R^{2}$) or low adsorption rates ($k R \ll 1$), as shown in
Fig. \ref{fig:semireflective} below.

In order to justify the form of the radiative
B.C. (\ref{eq:bcreflective}), we determine the backward B.C. on the
probability distribution from a well-known forward B.C.  For the
process defined in Sec. \ref{sec:model}, the forward B.C. equation on
the probability distribution is \cite{berezhkovskii:054115}
\begin{align}
D_2 \pd{p(r,\theta)}{r}_{\lvert r=R} = - \; k D_2 \; p(R,\theta) + \lambda p(\theta) \label{eq:forwardbc}.
\end{align}

Since the stochastic process under
study is Markovian, one obtains the following Chapman-Kolmogorov equation on the
conditional probabilities, for $\bar{t}>s>t'$,
\begin{align} 
 p(\bm{\bar{x}},\bar{t}|\bm{x}',t') &= \int^{\pi}_{0} \int^{R}_{0} 2 \pi r^2 dr \sin \theta d\theta \ \bar{p}(r,\theta) p(r,\theta) \nonumber \\ 
&+ \int^{\pi}_{0} 2 R^{2} \pi \sin \theta d\theta \ \bar{p}(\theta) p(\theta) .
\end{align}
Taking the derivative with respect to the intermediate time $s$ of the
above relation leads to the identity
\begin{align} \label{eq:chapman2}
&\int^{\pi}_{0}\int^{R}_{0} 2 \pi r^2 \sin \theta d\theta dr \left( \pd{p(r,\theta)}{s} \bar{p}(r,\theta)  
+ p(r,\theta)  \pd{\bar{p}(r,\theta)}{s} \right) \nonumber \\ 
&+ \int^{\pi}_{0} 2 \pi R^{2} \sin \theta d\theta \left( \pd{p(\theta)}{s} \bar{p}(\theta) + \pd{\bar{p}(\theta)}{s} p(\theta) \right) = 0. \nonumber
\end{align}
The next step is to substitute diffusion Eqs. (\ref{eq:forwarddiffb}
-- \ref{eq:Chapfrom2}) into the last relation.  The two terms
with the angular Laplace operators cancel each other due to its
hermiticity:
\begin{equation*}
\int^{\pi}_{0} \sin \theta d\theta \left( \Delta_{\theta} \ \bar{p}(\theta) p(\theta) - 
\Delta_{\theta} \ p(\theta) \bar{p}(\theta) \right) = 0 .
\end{equation*}
The divergence theorem applied on the bulk Laplacian
$\Delta_{(r,\theta)}$ and the backward equations
(\ref{eq:Chapfrom2},\ref{eq:Chapfrom1}) yield the
following relation over the sphere surface:
\begin{align*}
&  0 = \int^{\pi}_{0} 2 R^{2} \pi \sin \theta d\theta \left( \ D_{2} \pd{p(r,\theta)}{r}_{\lvert R} \bar{p}(r,\theta) \right. \\
& 		- D_{2} \pd{\bar{p}(r,\theta)}{r}_{\lvert R} p(r,\theta) + \lambda  \biggl\{\bar{p}(\theta) - \bar{p}(R,\theta)\biggr\} p(\theta) \\
& \left.	+ \biggl\{ -\lambda \ p(\theta) + k D_2 \ p(R,\theta)\biggr\} \bar{p}(\theta) \right),
\end{align*}
which is satisfied only if:
\begin{align} 
\label{eq:boundary}
 D_{2} & \pd{\bar{p}(r,\theta)}{r}_{\lvert R} \; p(R,\theta) = D_{2} \pd{p(r,\theta)}{r}_{\lvert R} \bar{p}(R,\theta)  \nonumber \\
 & - \lambda \ \bar{p}(R,\theta) p(\theta) + k D_{2} \ p(R,\theta) \bar{p}(\theta) .
\end{align}
Insertion  of the forward B.C. (\ref{eq:forwardbc}) into
Eq. (\ref{eq:boundary}) gives the B.C. on the backward probability
distribution
\begin{equation} \label{eq:backwardbc}
 \pd{p(\bm{\bar{x}},\bar{t}|(r,\theta),s)}{r}_{\lvert r= R} = k \bigl\{ p(\bm{\bar{x}},\bar{t}|\theta,s) - p(\bm{\bar{x}},\bar{t}|(r,\theta),s)\bigr\}_{\lvert R}.
\end{equation}
We then integrate Eq. (\ref{eq:backwardbc})
over the space and time variables $\bm{x}$ and $t$ according to Eqs. (\ref{def:t1}) and (\ref{def:t2})
to obtain the B.C. on the MFPT:
\begin{equation*}
\pd{t_{2}}{r}_{\lvert {\bf{r}}= (R,\theta)} = k \bigl\{ t_{1}(\theta) - t_{2}(R,\theta) \bigr\} \qquad (\epsilon \leq \theta \leq \pi),
\end{equation*}
which identifies with Eq. (\ref{eq:bcreflective}). Agreement with the B.C.
used in \cite{Rupprecht:2012a} is discussed in Appendix \ref{sec:boundary}.

\subsection{Integral equation}   \label{sec:integral_eq}

From the set of Eqs. 
(\ref{eq:temps1}) - (\ref{eq:bcreflective}) we now derive an integral equation on $t_1$ only. \\

We first recall that the eigenfunctions of the angular 
Laplace operator $\Delta_{\theta}$ of Eq. (\ref{eq:diffusionoperators3D})
are expressed in terms of the Legendre polynomials $P_n$:
\begin{equation}
\label{eq:Vn_def}  
- \Delta_{\theta} P_{n}( \cos \theta ) = \rho_n P_{n}(\cos \theta )   \quad  (n\geq 0),
\end{equation}
where $\rho_n = n(n+1)$.  We set $V_{n}(\theta) = \sqrt{2n+1}
P_{n}(\cos \theta)$ to get the orthonormality: $\langle V_{n} | V_{m}
\rangle = \delta_{nm}$, where the inner product is defined as
\begin{equation} 
\label{eq:scalrproduct}
(f,g) \rightarrow \langle f | g \rangle_{\epsilon} \equiv \frac{1}{2} \int^{\pi}_{\epsilon} f(\theta) g(\theta) \sin \theta d\theta,
\end{equation}
and $\langle f | g \rangle$ is the scalar product for $\epsilon = 0$.
We also define 
\begin{equation} \label{def:Imn}
K^{(\epsilon)}_{mn} \equiv \langle V_{n} | V_{m} \rangle_{\epsilon} \qquad (m , n \geq 0),
\end{equation}
with the explicit expressions listed in Table \ref{tab:Vtable}. \\

\begin{table} \label{}
 \begin{center}
 \begin{tabular}{|c|c|}
  \hline
$V_{n}(\theta)$ & $ \sqrt{2n+1} ~P_{n}(\cos \theta)$\\
  \hline
$u$ & $ \cos \epsilon$\\
  \hline
  $F_n(u), n \geq 1$   & $\sum\limits_{k=1}^n 2(u-1)P_k^2(u) + [P_k(u) - P_{k-1}(u)]^2 $\\
& $- (u-1)P_n^2(u) + (u-1)P_0^2(u) + u$ \\
  \hline
$g_{\epsilon}(\theta)$ & $\ln\left(\frac{1-\cos(\theta)}{1-\cos(\epsilon)}\right)$\\
  \hline
$\langle g_{\epsilon}|1 \rangle_{\epsilon} \equiv \langle g_{\epsilon} \rangle_{\epsilon} $ & 
$\ln\left(\frac{2}{1-\cos \epsilon}\right) - \frac{1+\cos \epsilon}{2}$\\
  \hline
$\xi_{n}~ (n \geq 1)$ & 
$- \frac{\sqrt{2n+1}}{2} \left( \left(1+ \frac{n u}{n+1} \right) P_{n}(u) + \frac{ P_{n-1}(u)}{n+1} \right)$ \\
  \hline
$K^{(\epsilon)}_{00}$ & $ \frac{1+\cos(\epsilon)}{2} $\\
  \hline
$K^{(\epsilon)}_{n0}$ ~ $(n\geq 1)$  & $ \frac{P_{n+1}(u) - P_{n-1}(u)}{2 \sqrt{2 n +1}} $\\
  \hline
$K^{(\epsilon)}_{nn}$ ~ $(n\geq 1)$  & 
$ \frac{F_{n}(u)+1}{2} $\\
 \hline
$K^{(\epsilon)}_{nm}$ ~ $(m \neq n)$ & $ \frac{\sqrt{2n+1}\sqrt{2m+1}}{2 \left( m(m+1) - n(n+1) \right)} \left( (m-n)u P_m(u) P_n(u) + \right.$ \\
$n,m \geq 1$ & $\left. + n P_{n-1}(u) P_m(u) - m P_{m-1}(u) P_n(u) \right) $ \\
  \hline
$I^{(\epsilon)}_{nn}$ ~ $(n\geq 1)$ & 
$\frac{2n+1}{2} \left( - P_n(u) \frac{u P_n(u) - P_{n-1}(u)}{n+1} + \frac{F_n(u) + 1}{2n+1} \right)$ \\
  \hline
$I^{(\epsilon)}_{nm}$ ~ $(m \neq n) $ & 
$\frac{\sqrt{2n+1}\sqrt{2m+1} m}{2 (n+1)[m(m+1)-n(n+1)]} \left((n-m) u P_m(u) P_n(u) \right.$ \\
$ m,n\geq 1$ & $ + (m+1)P_m(u) P_{n-1}(u)$\\
& $\left.- (n+1)P_n(u)P_{m-1}(u) \right)$\\
\hline
\end{tabular}
 \end{center}
\caption{
Summary of the quantities involved in the computation of the vector
$\bm{\xi}$ and the matrices $Q$ and $P$ in Eqs. (\ref{eq:defU}) and (\ref{eq:defQ}) that determine the Fourier coefficients $d_n$ of
$t_1(\theta)$ according to Eq. (\ref{eq:dn}).}
\label{tab:Vtable}
\end{table}

The starting point for solution of the set of equations (\ref{eq:temps1} and
(\ref{eq:temps2}) is a Fourier decomposition of $t_2(r,\theta)$,
\begin{equation} 
\label{eq:t2_def_particular} 
t_2(r,\theta) = \alpha_{0}- \frac{r^{2}}{2d D_2} + \sum^{\infty}_{n=1}
\alpha_{n} \left(\frac{r}{R}\right)^{n} V_{n}(\theta),
\end{equation} 
where $d = 3$ for the three-dimensional spherical cavity considered
here (in Appendix \ref{sec:disk}, we show how this approach can be
directly translated for two-dimensional problems).

Due to the B.C. (\ref{eq:bcreflective}), the projection
of $t_2(R,\theta)$ onto the orthonormal basis $\{V_n(\theta)\}_{n\geq
0}$, is
\begin{align} \label{eq:inteq}
\int^{\pi}_{0} t_2(R,\theta) V_{n}(\theta) \sin \theta d\theta &= \int^{\pi}_{\epsilon} t_1(\theta) V_{n}(\theta) \sin \theta d\theta \nonumber \\
 &- \int^{\pi}_{\epsilon} \frac{1}{k} \pd{t_2}{r}_{\lvert R} V_{n}(\theta) \sin \theta d\theta.
\end{align}
The Fourier decomposition (\ref{eq:t2_def_particular}) leads to linear
equations on the coefficients $\alpha_{0}$ and $\alpha_{n}$, $n \geq
1$,
\begin{align}
 \alpha_{0}- \frac{R^{2}}{2d D_2} \left(1 + \frac{2 K^{(\epsilon)}_{00}}{k R}\right) &= 
\braket{t_1}{1} -  \sum^{\infty}_{m=1} \frac{m K^{(\epsilon)}_{0m}}{k R} \ \alpha_{m},  \label{eq:sysalpha0_particular}\\ 
\alpha_{n} + \sum^{\infty}_{m=1} \frac{m K^{(\epsilon)}_{nm}}{k R} \ \alpha_{m} &= 
\braket{t_1}{V_{n}} + \frac{R K^{(\epsilon)}_{n0}}{d k D_2} .  \label{eq:sysalphan_particular}
\end{align}
First, one can solve the set of linear equations
(\ref{eq:sysalphan_particular}), independently of
Eq. (\ref{eq:sysalpha0_particular}), by writing
\begin{eqnarray}
\sum^{\infty}_{m=1} \left(\delta_{n,m} + M_{nm} \right) \alpha_{m} = \hat{U}_{n},
\end{eqnarray}
where
\begin{eqnarray} \label{def:QandU}
M_{nm}  &\equiv& \frac{m}{k R} K^{(\epsilon)}_{nm} , \\  \label{eq:Uhat}
\hat{U}_{n} &\equiv& \braket{t_1}{V_{n}} + \frac{R}{d k D_2} K^{(\epsilon)}_{n0}
\end{eqnarray}
for all $n,m \geq 1$.  Formally, the solution of this system of
equations is
\begin{eqnarray} \label{eq:alphanformal}
\alpha_{n} = \left[\left(\mathbb{I} + M \right)^{-1} \hat{U}\right]_{n},
\end{eqnarray}
where $(\mathbb{I})_{n,m} = \delta_{n,m}$ stands for the identity
matrix.  Note that $t_1(\theta)$ and thus $\hat{U}_n$ are still
unknown.

Second, substituting the B.C. (\ref{eq:bcreflective})
into the diffusion equation (\ref{eq:temps1}) leads to
\begin{equation} \label{eq:diffts}
 - \frac{D_1}{R^{2}}\Delta_{\theta} t_1(\theta) = -1 + \frac{\lambda}{k} \pd{t_2}{r}_{\lvert {\bf{r}}=(R,\theta)}  .
\end{equation}
The substitution of Eq. (\ref{eq:t2_def_particular}) into this
relation yields
\begin{equation}  \label{eq:diffts2}
 - \Delta_{\theta} t_1(\theta) = -\omega^{2} T
 + \omega^{2} \sum^{\infty}_{n=1} \alpha_{n} \frac{n}{k R}  V_{n}(\theta),
\end{equation}
where we defined
\begin{eqnarray} 
   \omega &\equiv& R\sqrt{\lambda/D_1}, \label{def:omega} \\
   T      &\equiv& \frac{1}{\lambda} + \frac{1}{\alpha}, \label{def:T}
\end{eqnarray}
$\alpha$ being the inverse of the mean re-adsorption time on the
surface after a desorption event, as defined in
\cite{berezhkovskii:054115}
\begin{equation} \label{def:alpha}
    \alpha \equiv \frac{d k D_2}{R}.
\end{equation}

The solution of Eq. (\ref{eq:diffts2}) which satisfies the
B.C. $t_1(\epsilon) =0$, is
\begin{equation} \label{eq:t1distribution2D}
t_1(\theta) = \omega^{2} T g_{\epsilon}(\theta) - \omega^2 \sum^{\infty}_{n=1} \alpha_{n} \frac{n}{k R}  
\frac{V_{n}(\theta)-V_{n}(\epsilon)}{\rho_n} 
\end{equation}
for $\epsilon < \theta < \pi$, and $t_1(\theta) = 0$ otherwise ($0
\leq \theta \leq \epsilon$).  Here, $R^2 g_{\epsilon}(\theta)/D_1$ is
the well-known MFPT for a surface search (i.e. when $\lambda = 0$)
\cite{Redner:2001a}:
\begin{equation}
\label{eq:g_eps}
g_{\epsilon}(\theta) = \ln\left(\frac{1-\cos(\theta)}{1-\cos(\epsilon)}\right) 
\end{equation}
for $\epsilon < \theta < \pi$, and $g_\epsilon(\theta) = 0$ otherwise.
One can easily check that $-\Delta_\theta g_\epsilon = -1$ and
$g_\epsilon(\epsilon) = 0$.
Using the formal expression (\ref{eq:alphanformal}) for $\alpha_{n}$
and introducing the dimensionless function
\begin{equation}
\label{eq:psi_t1}
\psi(\theta) = \frac{t_1(\theta)}{\omega^{2}T} ,
\end{equation}
one can represent Eq. (\ref{eq:t1distribution2D}) as an integral
equation on $\psi(\theta)$:
\begin{eqnarray} \label{eq:t1distribution_dimless}
\psi(\theta) = g_{\epsilon}(\theta) 
&+& \sum\limits^{\infty}_{n,m=1} \frac{V_{n}(\theta)-V_{n}(\epsilon)}{\rho_n} ~X_{nm} \\  \nonumber
&& \times \biggl(\frac{R}{d k D_2 T} K^{(\epsilon)}_{0m} + \omega^2 \braket{\psi}{V_{m}} \biggr)  
\end{eqnarray}
for $\epsilon < \theta < \pi$, and $\psi(\theta) = 0$ otherwise.
Here, we introduced
\begin{equation} \label{def:xcoeff}
 X_{nm} \equiv -\frac{n}{kR} \left[\left(\mathbb{I} + M \right)^{-1}\right]_{nm}.
\end{equation}

\subsection{Exact solution} \label{sec:exactresolution}

Expanding the function $\psi(\theta) - g_\epsilon(\theta)$ on the
complete basis of functions $\{V_{n}(\theta)-V_{n}(\epsilon)\}_{n\geq
1}$,
\begin{equation} 
\label{psi}
\psi(\theta) = g_{\epsilon}(\theta) + \sum^{\infty}_{n=1} d_{n} \{V_{n}(\theta)-V_{n}(\epsilon)\}  \quad (\epsilon < \theta < \pi),
\end{equation}
one obtains a set of linear equations for the unknown coefficients
$\{d_{n}\}_{n \geq 1}$
\begin{eqnarray} 
\label{eq:relationdn} 
\sum^{\infty}_{n=1} && d_{n} \{ V_ {n}(\theta)-V_ {n}(\epsilon) \} \\  \nonumber
&& = \omega^2 \sum^{\infty}_{n=1} \left( U_{n} +  \sum^{\infty}_{l=1} Q_{nl} d_{l} \right) \{ V_ {n}(\theta)-V_ {n}(\epsilon) \},
\end{eqnarray} 
where we defined the vectors $\bm{U}$ and $\bm{\xi}$ by their $n$-th
coordinates $(n\geq 1)$
\begin{align}
\label{eq:defU}
U_{n} &\equiv  \frac{1}{\rho_n} \sum^{\infty}_{m=1} X_{nm} \left( \frac{\xi_{m}}{\rho_m} + \frac{K^{(\epsilon)}_{0m} R}{d k D_2 \omega^{2} T}  \right), \\
\xi_{n} &\equiv \rho_n \langle  g_{\epsilon}|V_{n}\rangle_{\epsilon} ,
\end{align}
and the matrices $Q$ and $I^{(\epsilon)}$ by their $n$-th row and
$l$-th column ($n,l \geq 1$)
\begin{eqnarray} 
\label{eq:defQ}
Q_{nl} &\equiv& \frac{1}{\rho_n} \sum^{\infty}_{m=1}  X_{nm} ~ I^{(\epsilon)}_{ml} , \\
I^{(\epsilon)}_{ml}  &\equiv& \langle V_{m}(\theta)|V_{l}(\theta)-V_{l}(\epsilon)\rangle_{\epsilon}.
\end{eqnarray}
As Eq. (\ref{eq:relationdn}) is satisfied for all $\theta \in
[\epsilon,\pi]$, the coefficients $d_n$ are found as
\begin{equation} \label{eq:dn}
d_{n} = \left[\omega^2 \left(\mathbb{I} - \omega^2 Q\right)^{-1} U\right]_{n} .
\end{equation}
Combining this relation with Eqs. (\ref{eq:g_eps}, \ref{eq:psi_t1},
\ref{psi}), one finally obtains an exact representation for the MFPT
\begin{equation}  \label{eq:t1distribution}
t_1(\theta) = \omega^2 T \left(g_\epsilon(\theta) + \sum\limits_{n=1}^{\infty} d_n \{ V_n(\theta)-V_n(\epsilon)\} \right)
\end{equation}
for $\epsilon < \theta \leq \pi$, and $t_1(\theta) = 0$ for $0\leq
\theta \leq \epsilon$.

Averaging $t_1(\theta)$ over the whole surface and using the relation
\begin{equation*} 
\langle P_{n}(\cos \theta)-P_{n}(\cos \epsilon) | 1 \rangle_{\epsilon}  
= \rho_{n} \langle P_{n}(\cos \theta)| g_{\epsilon}(\theta)\rangle_{\epsilon} = \xi_{n} ,
\end{equation*}
we obtain the exact formula for the surface GMFPT:
\begin{equation}
\label{eq:searchtime} 
 \langle t_1 \rangle = \omega^2 T \left( \langle g_{\epsilon} \rangle + \sum^{\infty}_{n=1} d_{n} \xi_{n} \right),
\end{equation}
where $\langle g_{\epsilon} \rangle$ is computed by integrating
Eq. (\ref{eq:g_eps}):
\begin{equation}
 \langle g_{\epsilon} \rangle = \ln \left( \frac{2}{1 - \cos \epsilon} \right) + \frac{1 + \cos \epsilon}{2} .
\end{equation}
Equations (\ref{eq:t1distribution}) and (\ref{eq:searchtime}) are among the
main results of the paper, and several comments are in order:

(i) As expected, in both limits $\lambda = 0$ and $k = \infty$, we
retrieve the limit of the MFPTs for the surface search process alone:
$t_1(\theta) \to \frac{R^2}{D_1} g_{\epsilon}(\theta)$.

(ii) A physical interpretation of Eq. (\ref{eq:searchtime}) is that
the surface GMFPT is the product of the mean time $T = \lambda^{-1} +
\alpha^{-1}$ for an elementary cycle composed of one surface
exploration and one bulk excursion, by the mean number of cycles
before reaching the target.

(iii) A numerical implementation of the exact solutions in
Eqs. (\ref{eq:t1distribution}) and  (\ref{eq:searchtime}) requires a
truncation of the infinite-dimensional matrix $Q$ to a finite size
$N\times N$.  After a direct numerical inversion of the truncated
matrices $(\mathbb{I}+M)$ in Eq. (\ref{eq:alphanformal}) and
$(\mathbb{I}-\omega^2 Q)$ in Eq. (\ref{eq:dn}), the MFPTs from
Eqs. (\ref{eq:t1distribution}, \ref{eq:searchtime}) are approximated
by truncated series (with $N$ terms).  At a fixed tolerance threshold,
higher values of $\lambda$ require higher values of $N$.  In spite of
the truncation, we will refer to the results obtained by this
numerical procedure as {\it exact solutions}, as their accuracy can be
arbitrarily improved by increasing the truncation size $N$.

(iv) The expression (\ref{eq:searchtime}) is valid for arbitrary
target size $\epsilon$, provided that the series are truncated at
sufficiently high $N$.

Substituting Eqs. (\ref{eq:psi_t1}) and (\ref{psi}) into
Eq. (\ref{eq:Uhat}), we deduce the Fourier coefficients $\alpha_n$ of
$t_2(r,\theta)$ for $n \geq 1$
\begin{eqnarray} \label{eq:alphan_final}
\alpha_{n} = \sum^{\infty}_{m=1} && \left(\left(\mathbb{I} + M \right)^{-1}\right)_{nm} \times \\
&& \left( \omega^2T \biggl[\frac{\xi_{m}}{\rho_{m}} + (I^{(\epsilon)} \bm{d})_{m} \biggr] + \frac{R  K^{(\epsilon)}_{m0}}{d k D_2}\right), \nonumber
\end{eqnarray}
while $\alpha_0$ is found from Eq. (\ref{eq:sysalpha0_particular}).
The coefficients $\alpha_n$ determine an exact representation
(\ref{eq:t2_def_particular}) of the MFPT $t_2(r,\theta)$.  Therefore one gets a complete exact solution of the problem for any starting
point.  In particular, the bulk GMFPT $\langle t_2\rangle$ averaged over
uniformly distributed starting points in the bulk, reads
\begin{equation}
\langle t_2\rangle \equiv \frac{2\pi}{4\pi R^3/3} \int\limits_0^R dr ~ r^2 \int\limits_0^\pi d\theta~ \sin\theta ~ t_2(r,\theta)
= \alpha_0 - \frac{3R^2}{10 d D_2} ,
\end{equation}
since the other terms from Eq. (\ref{eq:t2_def_particular}) vanish due
to the orthogonality of $V_n(\theta)$.  Substituting an expression for
$\alpha_0$, one gets
%
\begin{equation} \label{eq:bulkgmfpt}
\langle t_2\rangle = \langle t_1 \rangle + \frac{R^2}{2 d D_2} \left(\frac25 + \frac{2K_{00}^{(\epsilon)}}{kR}\right) - 
\sum\limits_{m,n=1}^\infty \frac{m K^{(\epsilon)}_{0m}}{kR} \alpha_{m}.
\end{equation}
Note that the coefficients $\alpha_n$ in Ref. \cite{Rupprecht:2012a}
were simply proportional to the coefficients $d_{n}$, up to the third
order in $\epsilon$.  The mixed boundary condition
(\ref{eq:bcreflective}) results in the more sophisticated expression
(\ref{eq:alphan_final}) for $\alpha_n$.

\subsection{Existence of an optimum} \label{sec:optimum}

Despite the prediction of \cite{berezhkovskii:054115} that the bulk and surface GMFPTs
are monotonic functions of $\lambda$, the exact solutions prove to
admit a minimum, as seen in Fig. \ref{fig:mfpt2}b on the example of the surface GMFPT.
In this section we focus on the surface GMFPT and we determine sufficient
conditions for this GMFPT to be an optimizable function of
$\lambda$, which are set by two requirements: (i) desorption events
should decrease the search time for small enough values of $\lambda$,
i.e. $\langle t_1\rangle < \langle t_1\rangle_{\lambda = 0} = t_s$;
(ii) the mean surface search time is lower than the mean bulk search
time, i.e. $t_s < t_b$. 

The first condition is fulfilled when the derivative of the surface GMFPT is
negative at $\lambda = 0$. We first rewrite the coefficients $U_{n}$ as
\begin{equation}
 U_{n} = Z_{n} + \frac{1}{d k R+ \lambda \frac{R^2}{D_2}} \frac{D_1}{D_2} W_{n} ,
\end{equation}
where
\begin{eqnarray}
 Z_{n} &\equiv& \frac{1}{\rho_n} \sum^{\infty}_{m=1}  X_{nm}~ \frac{\xi_{m}}{\rho_m} , \label{eq:V}  \\
 W_{n} &\equiv& \frac{1}{\rho_n} \sum^{\infty}_{m=1}  X_{nm} ~ K^{(\epsilon)}_{0m}.  \label{eq:W}
\end{eqnarray}  
The derivative of the GMFPT at $\lambda = 0$ is
\begin{equation}
\left(\pd{\langle t_1 \rangle}{\lambda}\right)_{\lambda=0} \hspace*{-1mm} = 
\frac{R^{4}}{D^{2}_{1}} \left(\frac{D_1}{d k R D_2} \biggl(\langle
g_{\epsilon}\rangle + (\bm{\xi} \cdotp \bm{W}) \biggr)  + (\bm{\xi} \cdotp \bm{Z}) \right).
\end{equation}
This derivative is negative provided that
\begin{equation} \label{eq:lowerbound}
 \frac{D_2}{D_1} \geq \left(\frac{D_2}{D_1}\right)_{{\rm low}}, \quad
\left(\frac{D_2}{D_1}\right)_{{\rm low}} = - \frac{1}{d k R} \ \frac{\langle
g_{\epsilon}\rangle + (\bm{\xi} \cdotp \bm{W})}{(\bm{\xi} \cdotp \bm{Z})} ,
\end{equation}
where we used the inequality $(\bm{\xi} \cdotp \bm{Z}) < 0$. 

The second bound (ii) is obtained when the surface search time $R^2
\langle g_{\epsilon} \rangle /D_1$ (at zero desorption rate) is lower
than the search time at infinite desorption rate.  Using the
first-order asymptotic expansion of Eqs. (\ref{eq:taus3D},
\ref{eq:taub3D}), this condition is explicitly given up to the second
order in $\epsilon \ll 1$ as
\begin{eqnarray} \label{eq:higherbound}
\frac{D_2}{D_1} &\leq& \left(\frac{D_2}{D_1}\right)_{{\rm up}}, \\\nonumber
\left(\frac{D_2}{D_1}\right)_{{\rm up}} &=& \frac{\pi}{3 \epsilon (1 - \epsilon \ln \epsilon) (2\ln(2/\epsilon)-1)} + O(\epsilon).
\end{eqnarray}
Combining the above inequalities, one gets a sufficient condition for
the surface GMFPT to be optimizable for $\epsilon \ll 1$:
\begin{equation}
\label{eq:optim_region_3d}
\left(\frac{D_2}{D_1}\right)_{{\rm low}} \leq \frac{D_2}{D_1} \leq \left(\frac{D_2}{D_1}\right)_{{\rm up}} .
\end{equation}
Figure \ref{fig:D2critiquevsk} displays the regime of parameters for
which the surface GMFPT is optimizable.

\subsection{Perturbative solution} \label{sec:approximate}

The first terms of a perturbative expansion with respect to $\epsilon$
of Eq. (\ref{eq:searchtime}) can easily be obtained.  At first order
in $\epsilon$, one has
\begin{align*}
\begin{split}
U_n & = U_n^{(0)} + O(\epsilon) = \frac{\sqrt{2n+1}}{n^2(n+1)^2} \frac{\frac{n}{kR}}{1+\frac{n}{kR}} + O(\epsilon),  \\
Q_{mn} & = Q_{mn}^{(0)} + O(\epsilon) = \frac{1}{n(n+1)}  \frac{\frac{n}{kR}}{1+\frac{n}{kR}} \delta_{m,n} + O(\epsilon), \\
\end{split}
\end{align*}
from which
\begin{align*}
d_n &=\omega^2 \bigl[(\mathbb{I} - \omega^2 Q^{(0)})^{-1} U^{(0)}\bigr]_n + O(\epsilon) \nonumber \\
    &= \frac{\omega^{2}}{n(n+1)} ~ \frac{\left(\frac{\frac{n}{kR}}{1+\frac{n}{kR}}\right)(2n+1)}{n(n+1) 
    + \omega^{2} \left(\frac{n/kR}{1+n/kR}\right)} + O(\epsilon) .
\end{align*}
One finds therefore
\begin{align}
\label{eq:psi_3d_perturb}
&\psi(\theta) = -2\ln\epsilon +2\ln\left(2\sin(\theta/2)\right) \nonumber \\ & - \omega^{2}\sum_{n=1}^\infty 
\left(\frac{\frac{n}{kR}}{1+\frac{n}{kR}}\right)\frac{2n+1}{n(n+1)}~\frac{1-P_n(\cos\theta)}{n(n+1)+ \omega^{2}\left(\frac{\frac{n}{kR}}{1+\frac{n}{kR}}\right)} 
\nonumber \\& +O(\epsilon).
\end{align}
Averaging the latter relation over $\theta$ yields:
\begin{align}
\label{eq:t1_perturb}
&\langle t_1 \rangle = \omega^2 T \biggl( 2\ln (2/\epsilon) - 1 \biggr. \nonumber \\ & 
- \left. \omega^{2} \sum_{n=1}^\infty \frac{2n+1}{n(n+1)}~\frac{\frac{n}{kR}}{n(n+1)(1+\frac{n}{kR})+ \omega^{2} \frac{n}{kR}}+O(\epsilon) \right).
\end{align}

Notice that this expression is identical to the perturbative
development at $\epsilon \ll 1$ for the case of a uniformly
semi-reflecting sphere, {\it including the target}
\cite{Rupprecht:2012a}.  In the limit of very small $\epsilon$, the
target is mainly reached from the adsorbed (surface) state, while its
reactivity from the bulk is expected to be negligible.  As seen on
Fig. \ref{fig:mfpt2}, the smaller the desorption rate $\lambda$, the
larger the domain of applicability in $\epsilon$ of the perturbative
development.  Note also that in Fig. \ref{fig:mfpt2}b, the first-order
expression exhibits a minimum with respect to the desorption rate.

\section{Comparison with the coarse-grained method} \label{sec:comparison}

\subsection{Coarse-grained approach}

In the coarse-grained approach of \cite{berezhkovskii:054115}, the
bulk, the target and the rest of the surface are considered as
effective states, denoted by $b$, $s$ and $\varnothing$, with no inner
spatial degrees of freedom.  An effective set of kinetic equations
combines four first-order reaction rates: (i) the rate $\lambda$
(denoted as $\beta$ in \cite{berezhkovskii:054115}) which is associated with
the desorption ($s \to b$); (ii) the rate $\alpha$ defined in
Eq. (\ref{def:alpha}) as the inverse of the mean re-adsorption time on
the sphere, which is associated with the effective adsorption ($b \to s$); (iii) the rate $k_s$ defined as the inverse of the surface GMFPT of
Eq. (\ref{eq:taus3D}) for the surface search alone ($s \to
\varnothing$): 
\begin{equation}
k_s = \frac{A}{t_{s}}, 
\end{equation}
where the prefactor $A$ from Eq. (\ref{eq:A}) accounts for the
difference between the averages over the initial position mentioned
above; (iv) the rate $k_b$ defined as the inverse of the bulk GMFPT of
Eq. (\ref{eq:taub3D}) for the bulk search alone ($b \to \varnothing$).
As in \cite{berezhkovskii:054115}, we define 
\begin{equation}
k_b \equiv  \frac{d D_2 \epsilon}{\pi R^2} , 
\end{equation}
which is the inverse of the first-order asymptotics in the limit
$\epsilon \ll 1$ of Eq. (\ref{eq:taub3D}).  


In the following, we focus on the MFPT averaged over the sphere
surface $\langle t_1 \rangle$ and compare our exact expression
Eq. (\ref{eq:searchtime}) to the concise approximate expression
derived in \cite{berezhkovskii:054115} for the surface GMFPT
\begin{equation} \label{eq:coarsegrained}
 \langle t_1 \rangle \simeq A\frac{\alpha  + \lambda + k_{b}}{\alpha k_{s} + \lambda k_{b} + k_{b} k_{s}}.
\end{equation}
As found in \cite{berezhkovskii:054115}, this expression predicts a
monotonic behavior for the surface GMFPT as a function of the desorption rate
$\lambda$.  Similar results can be obtained for the MFPT averaged over
the sphere volume $\langle t_2 \rangle$.

\subsection{Comparison}

\begin{figure}[h!]
 \centering
\includegraphics[width=90mm]{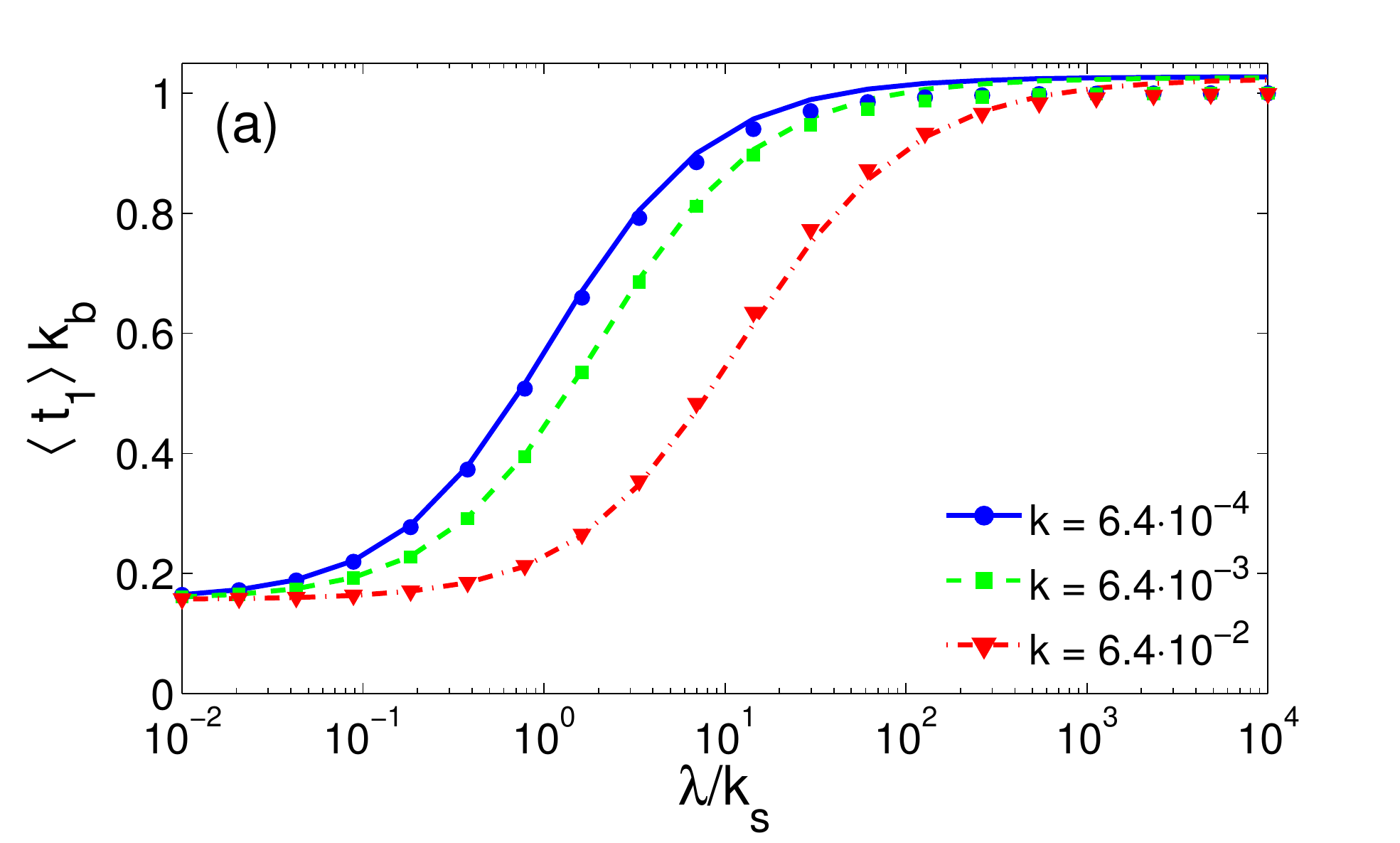}
\includegraphics[width=90mm]{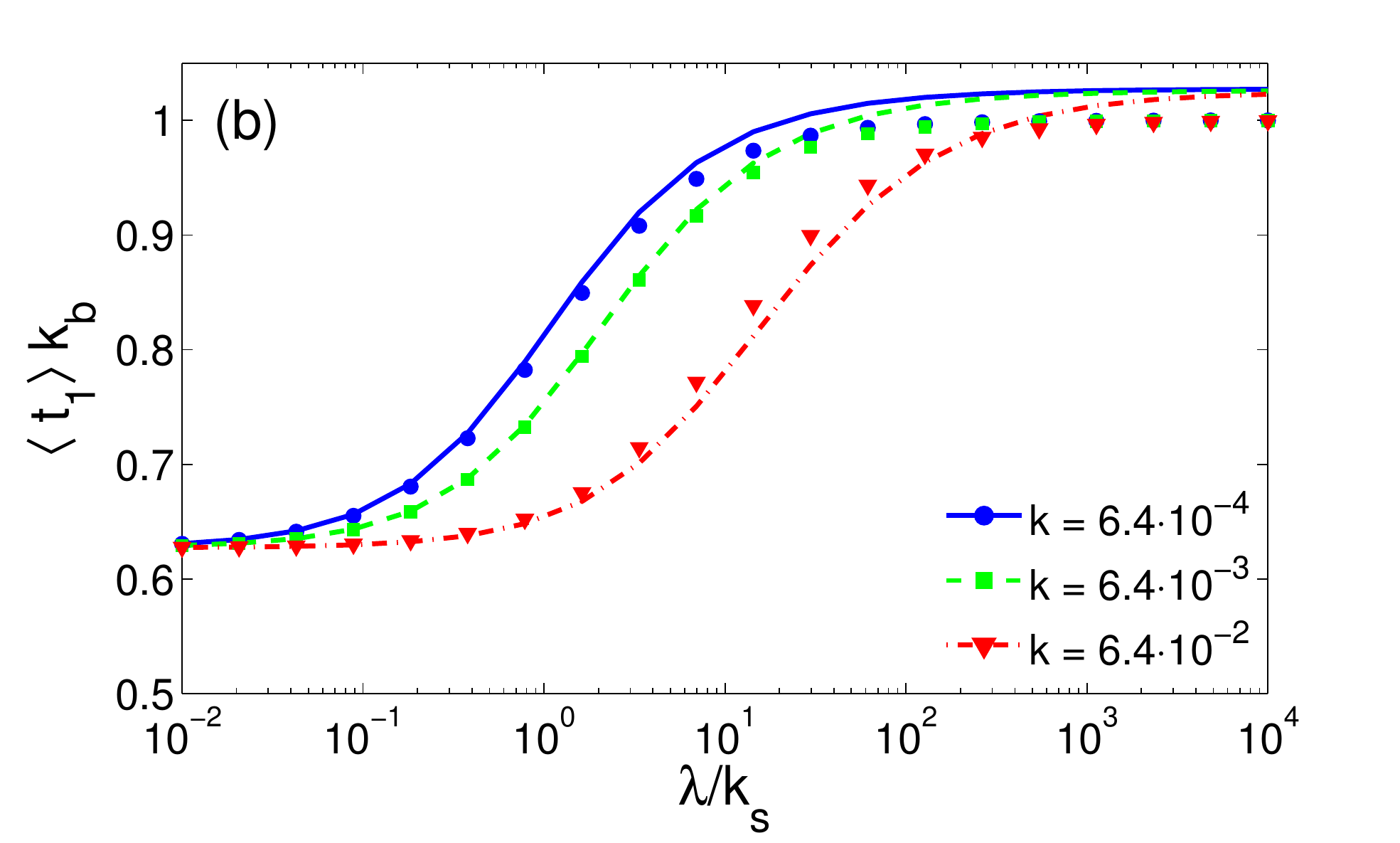}
\caption{
(Color online)  Comparison between the coarse-grained (symbols) and
exact (lines) surface GMFPT $\langle t_1 \rangle$ as a function of the
desorption rate $\lambda$, for $\epsilon = 0.02$, $N = 3 \cdot 10^{4}$
and several values of $kR = 6.4\cdot 10^{-4},\ 6.4\cdot 10^{-3},\
6.4\cdot 10^{-2}$ (from \cite{berezhkovskii:054115}), with $D_2 = D_1
= 1$ {\bf (a)} and $D_2 = 4 D_1 = 4$ {\bf (b)} in arbitrary units in
which $R = 1$.  }
\label{fig:mfpt1}
\end{figure}

\begin{figure}[h!]
 \centering
\includegraphics[width=90mm]{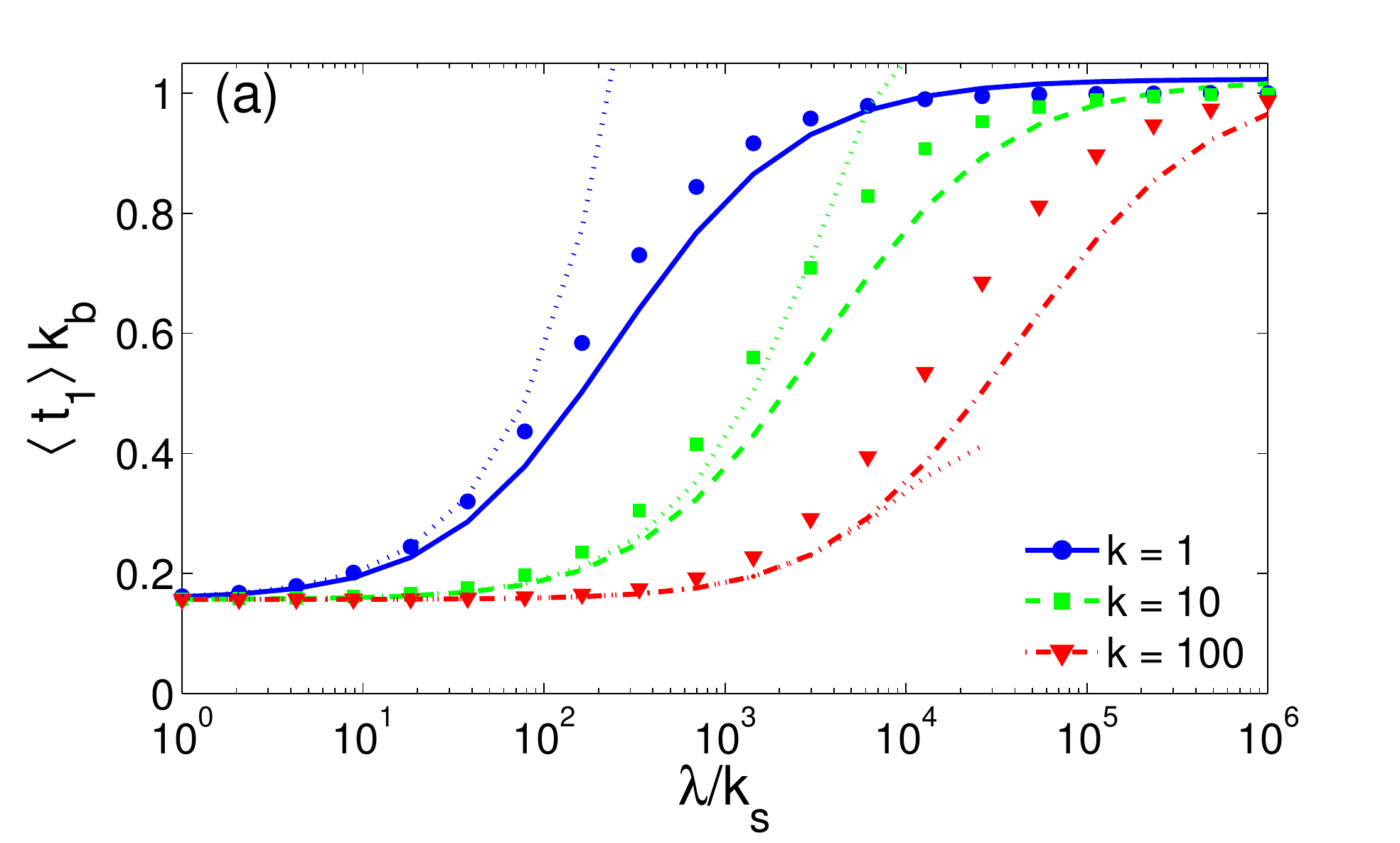}
\includegraphics[width=90mm]{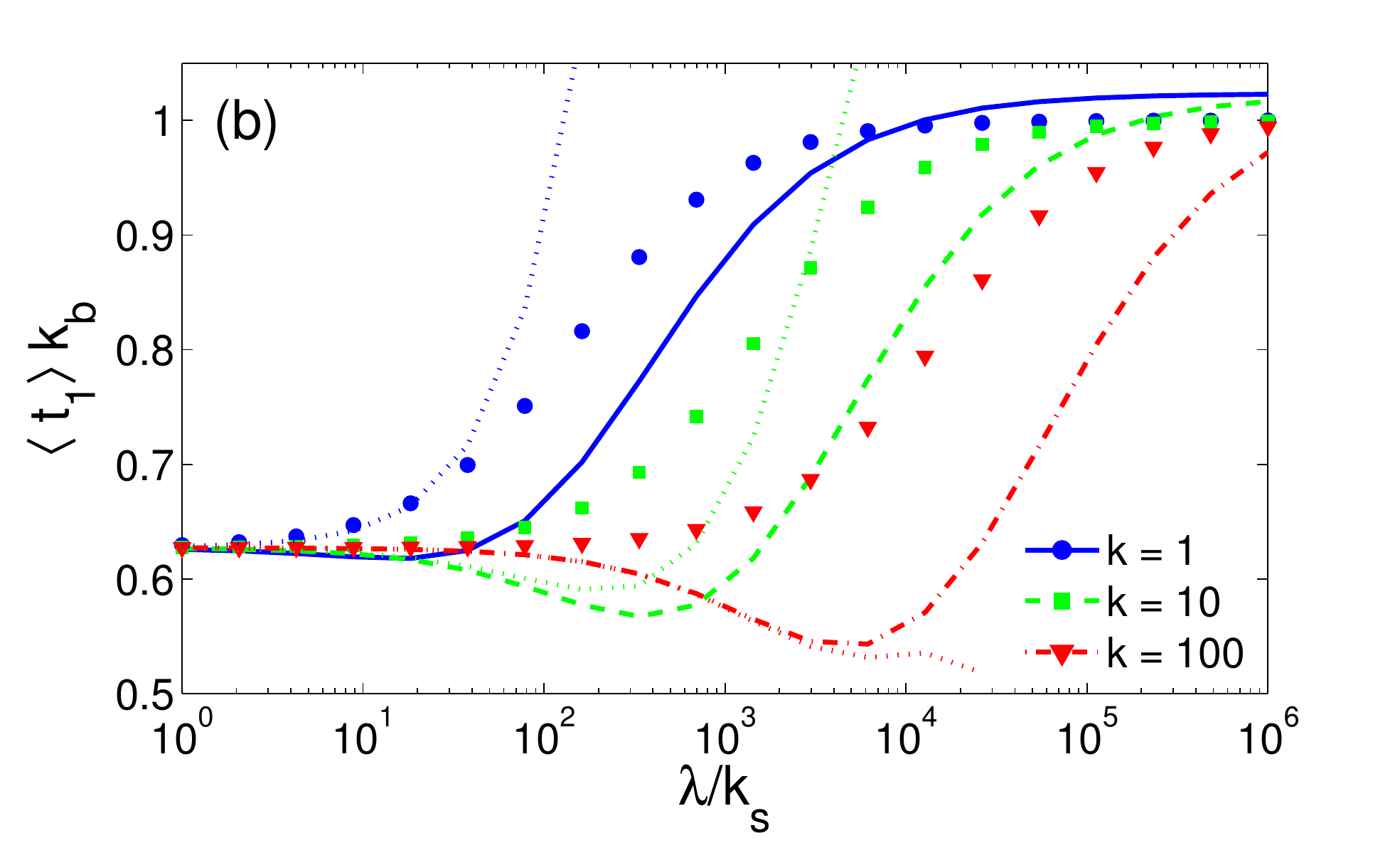}
\caption{
(Color online).  Comparison between the coarse-grained (symbols) and
exact (lines) surface GMFPT $\langle t_1 \rangle$ as a function of the
desorption rate $\lambda$, for $\epsilon = 0.02$, $N = 3 \cdot 10^{4}$
and several values of $kR = 1, 10, 100$, with $D_2 = D_1 = 1$ {\bf
(a)} and $D_2 = 4 D_1 = 4$ {\bf (b)}.  The dotted curves illustrate
the perturbative solution (\ref{eq:t1_perturb}) which is accurate for
moderate $\lambda$ but strongly deviates for very large $\lambda$.  }
\label{fig:mfpt2}
\end{figure}

Figure \ref{fig:mfpt1} shows that the coarse-grained approach and the
exact solution are in good agreement for the values of the adsorption
parameter $kR = 6.4\cdot 10^{-4},\ 6.4\cdot 10^{-3},\ 6.4\cdot
10^{-2}$, which are used for an analogous plot in
\cite{berezhkovskii:054115}.  However, as soon as $kR$ is large
enough, the exact and approximate curves are significantly different,
as illustrated on Fig. \ref{fig:mfpt2} (for $kR = 1, \ 10, \ 100$).
Finally, and most importantly, the exact solution for the surface GMFPT
$\langle t_1 \rangle$ can exhibit a minimum with respect to $\lambda$
(see Fig. \ref{fig:mfpt2}b) as opposed to the coarse-grained approach
which always predicts a monotonic behavior. Similarly, the bulk GMFPT
$\langle t_2 \rangle$ from Eq. (\ref{eq:bulkgmfpt}) also exhibits a
minimum with respect to $\lambda$ for the parameters used in
Fig. \ref{fig:mfpt2}b (not shown).

Note that in all considered cases of Figs. \ref{fig:mfpt1},
\ref{fig:mfpt2} the coarse-grained approach underestimates the search
time at large $\lambda$.  This discrepancy is due to the approximate
expression of $t_b$ used in \cite{berezhkovskii:054115} to estimate
$k_b$.  As expected, in a similar two-dimensional problem, our
solution at large $\lambda$ coincides with the earlier exact result of
\cite{Singer:2006b} for the mean exit time when the surface is
perfectly reflecting (see Appendix \ref{sec:disk}).

\begin{figure}[h!]
 \centering
\includegraphics[width=90mm]{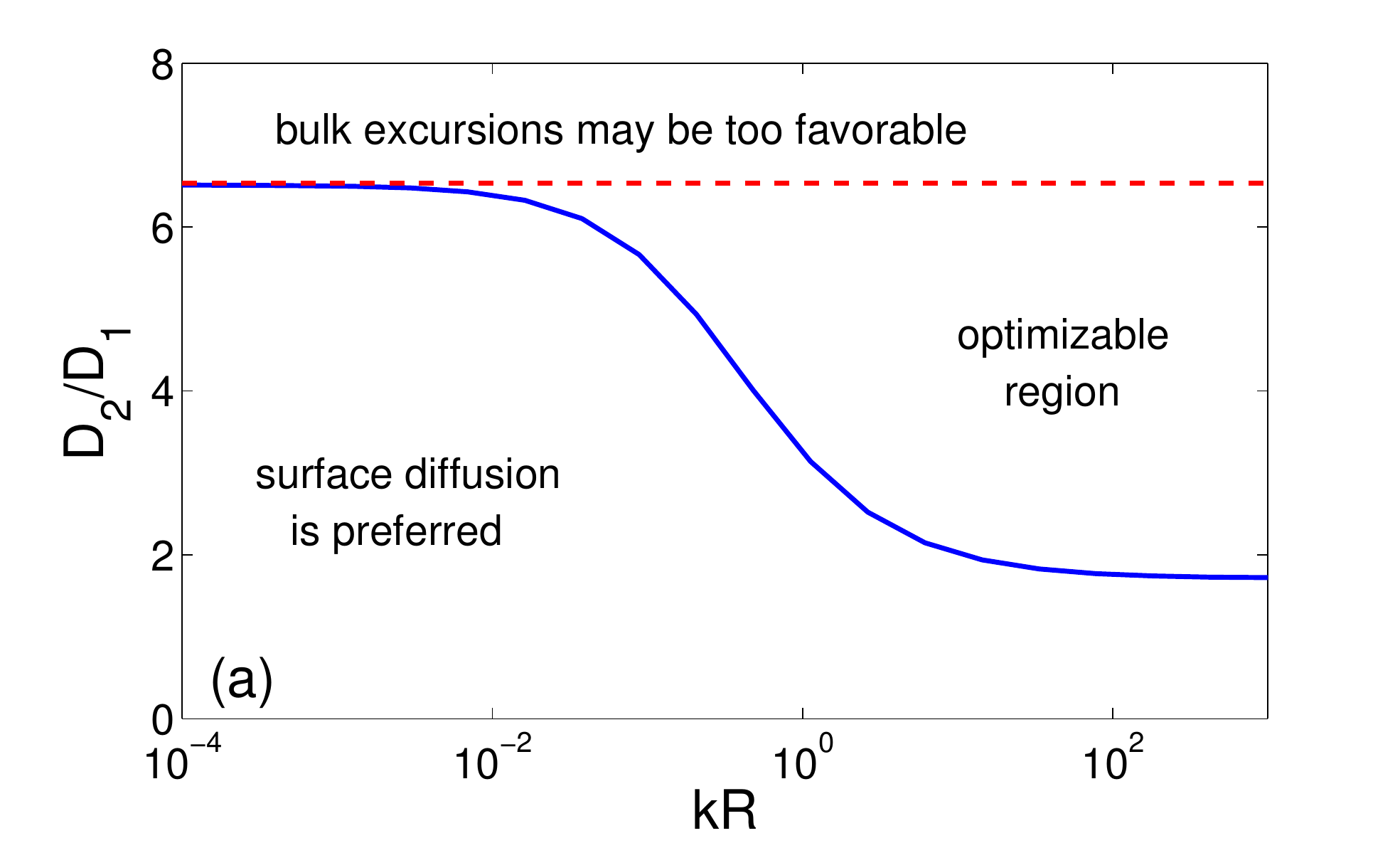}
\includegraphics[width=90mm]{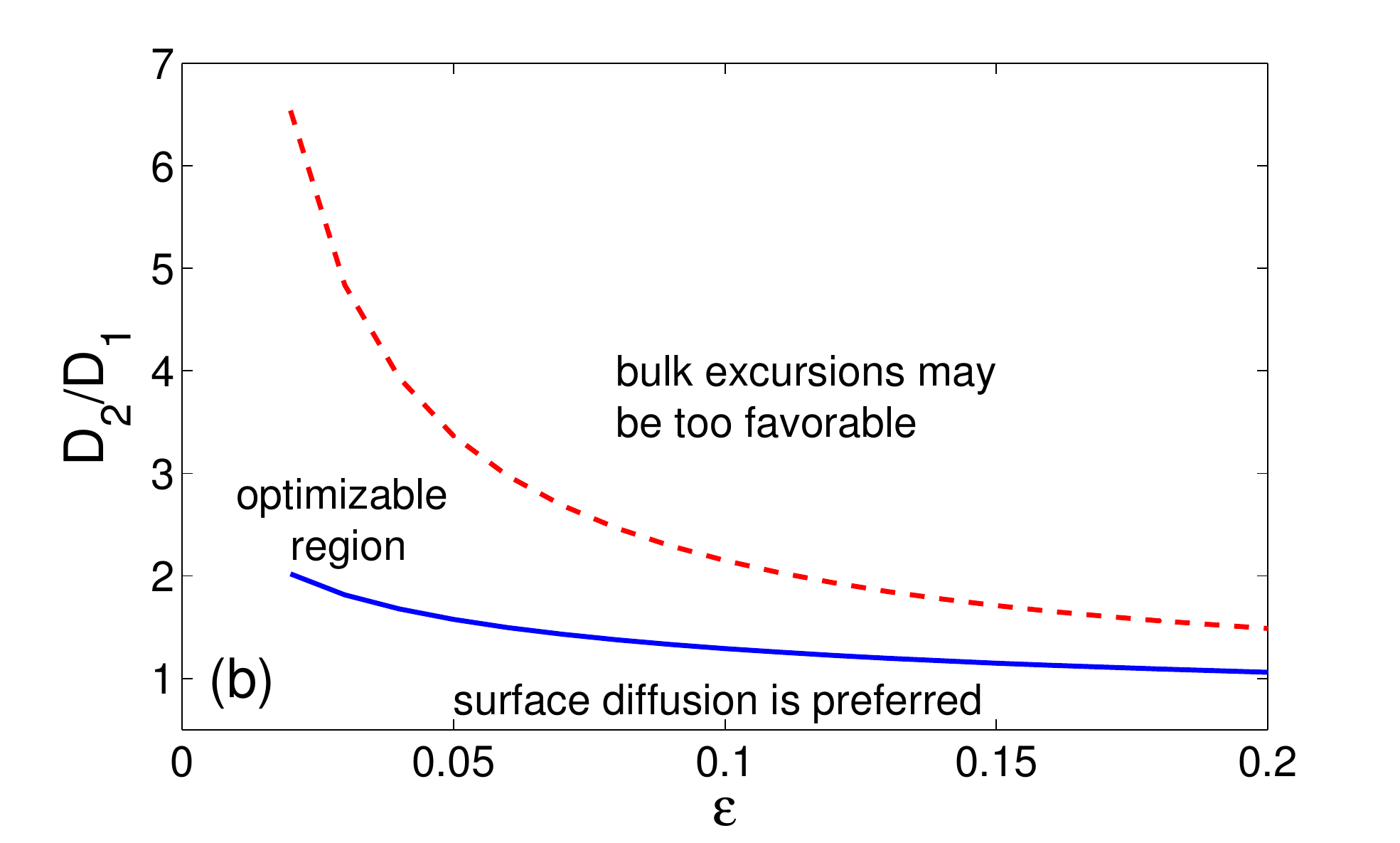}
\caption{
(Color online).  The regions of optimality for the surface GMFPT $\langle
t_1\rangle$.  {\bf (a)} The critical ratio $D_2/D_1$ as a function of
the adsorption coefficient $k$ for a fixed value of the target
half-width $\epsilon = 0.02$; {\bf (b)} The critical ratio $D_2/D_1$
as a function of the target size $\epsilon$, for the adsorption
coefficient $kR= 10$.  Below the lower bound (dashed red line),
surface diffusion is preferred.  Above the upper bound (solid blue
line), the GMFPT is smaller in the desorbed state than in the adsorbed
state.  In between, the surface GMFPT is an optimizable function of $\lambda$.
Series are truncated at $N = 10^{4}$.}
\label{fig:D2critiquevsk}
\end{figure}

\subsection{Discussion} \label{sec:physics}

We now discuss quantitatively the validity domain of the
coarse-grained approach and explain qualitatively why it fails to
reproduce the minimum of the surface GMFPTs with respect to $\lambda$.
 
As suggested in \cite{berezhkovskii:054115}, the validity of the
coarse-grained approach requires the equilibration times for
homogenization within each state to be faster than the other time
scales of the process.  In particular, the equilibration rate should
be larger than the following: \\
(i) the target encounter rate, i.e.,
\begin{align}
 \max(k_s,k_b) 					&\ll \min(D_1/R^2,D_2/R^2), 		\label{eq:rates1}
\end{align}
which implies $\epsilon\ll 1$.  The coarse-grained approach can only
describe narrow escape situations $\epsilon\ll 1$, as mentioned in
\cite{berezhkovskii:054115}, whereas the exact solution presented here
is valid for arbitrary $\epsilon$. \\
(ii) the inverse of the mean time for re-adsorption on the surface
after desorption, i.e.,
\begin{align} 
\alpha = 3 k D_2 / R      			&\ll \min(D_1/R^2,D_2/R^2), 		\label{eq:rates2}
\end{align}
which implies in particular that $kR \ll 1$.  Indeed, as seen in
Appendix \ref{sec:correlations}, in the regime $kR \ll 1$, the spatial
correlations between the starting and ending points of bulk excursions
are negligible.  The condition (\ref{eq:rates2}) is satisfied in the
situations displayed in Fig. \ref{fig:mfpt1}. 

However, Eq. (\ref{eq:rates2}) is not longer satisfied in both situations of
Fig. \ref{fig:mfpt2}, and one notices that the coarse-grained solution
does not match with the exact result.  This is particularly visible in
Fig. \ref{fig:mfpt2}b, where the surface GMFPT $\langle t_1 \rangle$ exhibits
a minimum with respect to the desorption rate $\lambda$.

The condition (\ref{eq:rates2}) for the applicability of the
coarse-grained approach turns out to be incompatible with the
existence of the minima of the bulk and surface GMFPTs with respect to $\lambda$.  Such a
minimum can be attributed to the fact that bulk excursions reduce the
time loss due to the recurrence of surface Brownian motion by bringing
the particle, through the bulk, to unvisited regions of the surface.
In the coarse-grained approach, a bulk excursion $s
\rightarrow b \rightarrow s$ ``consumes'' time but brings the particle
back to its initial effective state $s$.  The assumption that the
starting and ending points of bulk excursions belong to the same
effective state $s$ necessarily eludes the optimization property.

More precisely, bulk excursions are expected to be favorable if they
are fast (e.g. large adsorption rate $\alpha$) and long-ranged
($kR\ll1$, see Appendix \ref{sec:correlations}).  This is in clear
contradiction with the condition (\ref{eq:rates2}) of applicability of
the coarse-grained approach. Due to the relation $\alpha = 3 k D_2/R$,
this condition is achieved for a high enough diffusion coefficient
$D_2$, which sets the lower bound of $D_2/D_1$ derived above.
However, if the ratio $D_2/D_1$ is too large, bulk excursions can be
too favorable so that the optimum would be achieved for $\lambda =
\infty$.  The upper bound $\left(D_2/D_1\right)_{{\rm
up}}$ in Eq. (\ref{eq:optim_region_3d}) excludes this possibility.
The set of conditions on the diffusion coefficients ratio is
illustrated on Fig. \ref{fig:D2critiquevsk}.

Understanding by analytical means the
agreement between the exact and coarse-grained
expressions of Eqs. (\ref{eq:searchtime}) and (\ref{eq:coarsegrained}) 
under the assumptions of Eqs. (\ref{eq:rates1}) and (\ref{eq:rates2}) is a challenging question. 
A first step in this direction is presented in Appendix \ref{sec:agreement}, in which the 
agreement is found within the domain of applicability of the perturbative expansion 
(introduced in Sec. \ref{sec:approximate}).

\section{Conclusion}

We have obtained an exact expression for the MFPT and their spatial
averages, called bulk and surface GMFPTs, for the process studied in
\cite{berezhkovskii:054115}.  Compared to
\cite{Benichou:2011a,Benichou:2010}, the introduction of the surface
binding rate $k$ allowed one to avoid using the ejection distance $a$
(i.e. to set $a=0$) after each desorption events.  In contrast to the
statement of \cite{berezhkovskii:054115}, we have shown that the bulk
and surface GMFPTs can be optimized even in this situation and that
this optimality property is not related to the non-locality of the
intermittent process considered in
\cite{Benichou:2011a,Benichou:2010}.

These exact results can be extended in several directions to include the following:
(i) the search for a semi-reflecting target, with an adsorption
parameter different from the rest of the surface (see Appendix
\ref{sec:semi}); (ii) the 2D search for an angular aperture on the
boundary of a disk (see Appendix \ref{sec:disk}); and (iii) the biased
search for an arbitrary number of regularly spaced targets over an
otherwise semi-reflecting annulus (2D) or cone (3D), following the
method of \cite{Rupprecht:2012a}.  Notably, even for a 3D search for a
purely reflecting target and for a 2D search with a bulk diffusion
coefficient $D_2$ smaller than the surface diffusion coefficient
$D_1$, the surface GMFPT can be an optimizable function with respect to the
desorption rate (see Figs. \ref{fig:semireflective} and \ref{fig:mfpt2D} below).

\acknowledgments
O.B. is supported by the ERC Starting Grant No. FPTOpt-277998.

\appendix

\section{Boundary condition on the MFPT} \label{sec:boundary}

The diffusion equation and the boundary condition for the occupation
probability distribution in the bulk appear at first sight to be
different in \cite{berezhkovskii:054115} and \cite{Rupprecht:2012a}.
In this section, we show that these two sets of equations are in
agreement and lead to the same radiative boundary condition
(\ref{eq:bcreflective}) for the MFPT.

In \cite{Rupprecht:2012a} we defined a different set of equations for
the conditional probability distribution to include a radial ejection
distance $a$ after each desorption event, in the presence of a
velocity field.  With the shorthand notations of
Sec. \ref{eq:basiceq}, the forward advection-diffusion equation of
\cite{Rupprecht:2012a} reads:
\begin{align} 
 &\pd{p(r,\theta)}{r}_{\lvert R} = - \; k \; p(R,\theta) + \frac{v(R)}{D_{2}} p(R,\theta) \label{eq:bcaneq0},\\
 &\pd{ p(r,\theta)}{s} = D_2 \ \left(\Delta_{(r,\theta)} + \frac{v(r)}{D_{2} r}\right) \ p(r,\theta) \nonumber \\  
 & \ \ +   \lambda \ \left(\frac{R}{R-a}\right)^{2} \; \delta^{3}({\bf{r}}-(R-a,\theta)) \ p(\theta) , \label{eq:probdensity}
\end{align}
where $v(r)$ is a radial velocity field positive for an outward drift. 
The other diffusion equations on the conditional probabilities are the
same as in \cite{Rupprecht:2012a}.

Notice that Eq. (\ref{eq:bcaneq0}) does not involve the desorption
rate $\lambda$. However there is no contradiction with
Eq. (\ref{eq:forwardbc}), as Eq. (\ref{eq:bcaneq0}) (without drift)
leads the same backward boundary Eq. (\ref{eq:backwardbc}) as long the
appropriate limit for $a= 0$ in the Dirac function in
Eq. (\ref{eq:probdensity}) is
\begin{align}
\nonumber
 \int^{R}_{0} \hspace*{-1mm} \left(\frac{R}{R-a}\right)^{2} \hspace*{-1mm} &
 \delta(r-(R-a,\theta)) p(r,\theta) r^2 d r \xrightarrow{a \rightarrow 0}  R^2  p(R,\theta),
\end{align}
It can be proved that this condition is required from normalization of
the probability density.

\section{Analytical agreement between the coarse-grained
and exact solutions} \label{sec:agreement}

Figure \ref{fig:mfpt1} shows a good agreement between the exact and
coarse-grained expressions (\ref{eq:searchtime},
\ref{eq:coarsegrained}).  However, finding an explicit analytical
relation between these two expressions under the general conditions
(\ref{eq:rates1}, \ref{eq:rates2}) seems non-trivial.

We focus here on the following specific successive limits: (i) small
target extension ($\epsilon \ll 1$); (ii) low-desorption rate regime
($\lambda \ll k_s$); (iii) and intermediate range for the adsorption
rate: $k_b \ll \alpha \ll \min(D_2/R^{2},D_1/R^{2})$.

On one hand, the coarse-grained expression in these limits reads
\begin{align} \label{eq:limitcoarsegrained}
 \langle t_1 \rangle &\approx \left(1+\frac{\lambda}{\alpha}\right) t_s 
 + O\left(\epsilon,\frac{\lambda}{k_s}\right),
\end{align}
where we have used that $k_b/k_s \ll 1$ and $k_b/\alpha \ll 1$ for a
fixed value of $\alpha$ at sufficiently small $\epsilon \ll 1$.

On the other hand, in the limit $\epsilon \ll 1$ the perturbative
expansion of Eq. (\ref{eq:t1_perturb}) is accurate. In the
above-mentionned limits it reads
\begin{align} \label{eq:limitexact}
 \langle t_1 \rangle &\approx \left(1+\frac{\lambda}{\alpha}\right) t_s + O(\epsilon,\lambda/k_s,kR),
\end{align}
where we have used that
\begin{align}
&\sum_{n=1}^\infty \frac{2n+1}{n(n+1)}~\frac{\frac{n}{kR}}{n(n+1)(1+\frac{n}{kR})+ \frac{\lambda R^2}{D_1} \frac{n}{kR}} \nonumber \\
&\approx 1 + O(k R,R^2 \lambda/D_1),
\end{align}
and that the condition $R^2 \lambda/D_1 \ll 1$ is guaranteed from the
condition $\lambda/k_s \ll 1$.

The identification of the first order terms in
Eqs. (\ref{eq:limitcoarsegrained}, \ref{eq:limitexact}) shows
analytically the agreement between the coarse-grained and exact
solutions within a range of parameters which necessarily satisfies
Eqs. (\ref{eq:rates1}, \ref{eq:rates2}).  Notice that the argument
presented here relies on the condition $k_b / \alpha \ll 1$ which is
not satisfied in the situations represented on Fig. \ref{fig:mfpt1}.

\section{Measure of correlations} \label{sec:correlations}

In this section we quantify the spatial correlations between the
starting and ending points of a bulk excursion.  We then provide the
range of values for $k$ in which the spatial correlations are
negligible.  The probability density $\Pi(\theta)$ for a particle
initially started from the surface state at the angle $\theta_0 = 0, \
\phi_0 = 0$ to first return on the surface to any point
$(\theta,\phi)$ (with $\phi \in [0 , 2 \pi]$) is
\cite{Rupprecht:2012a}
\begin{equation}
 \Pi (\theta) = \frac{\sin \theta}{2} \left(1 + \sum^{\infty}_{n=1} \frac{2n+1}{1+\frac{n}{kR}}   
P_{n}(\cos \theta ) \right).
\end{equation}
The cumulative probability distribution for the relocation angle
$\theta \in [0,\pi]$ is the integral of the probability density
\begin{equation} \label{eq:probabilitydistribution}
 F(\theta) \equiv \frac{1}{2}\int^{\theta}_{-\pi} \Pi (\theta') \sin \theta'~ d\theta' .
\end{equation}
By analogy with the Kolmogorov-Smirnov test \cite{Massey:1951}, we
propose to measure the spatial correlations between the starting and
ending points of a bulk excursion by the norm
\begin{equation}
 N_K \equiv \max_{\theta \in [0,\pi]} \nm{\delta F(\theta)}  ,
\end{equation}
with
\begin{align} \label{eq:cumulativedifference}
\delta F(\theta) \equiv F(\theta) - F_{u}(\theta) =
\frac{1}{2} \sum^{\infty}_{n=1} \frac{P_{n-1}(\cos \theta) - P_{n+1}(\cos \theta)}{1+\frac{n}{kR}},
\end{align}
and $F_{u}(\theta) = (1-\cos \theta)/2$ is the cumulative distribution
for uncorrelated random relocation on the sphere.  This leads a
correlation angle $\Theta$ which is defined as the solution of the
equation $\delta F(\Theta) = N_K$.

\begin{figure}[h!]
\centering
\includegraphics[width=90mm]{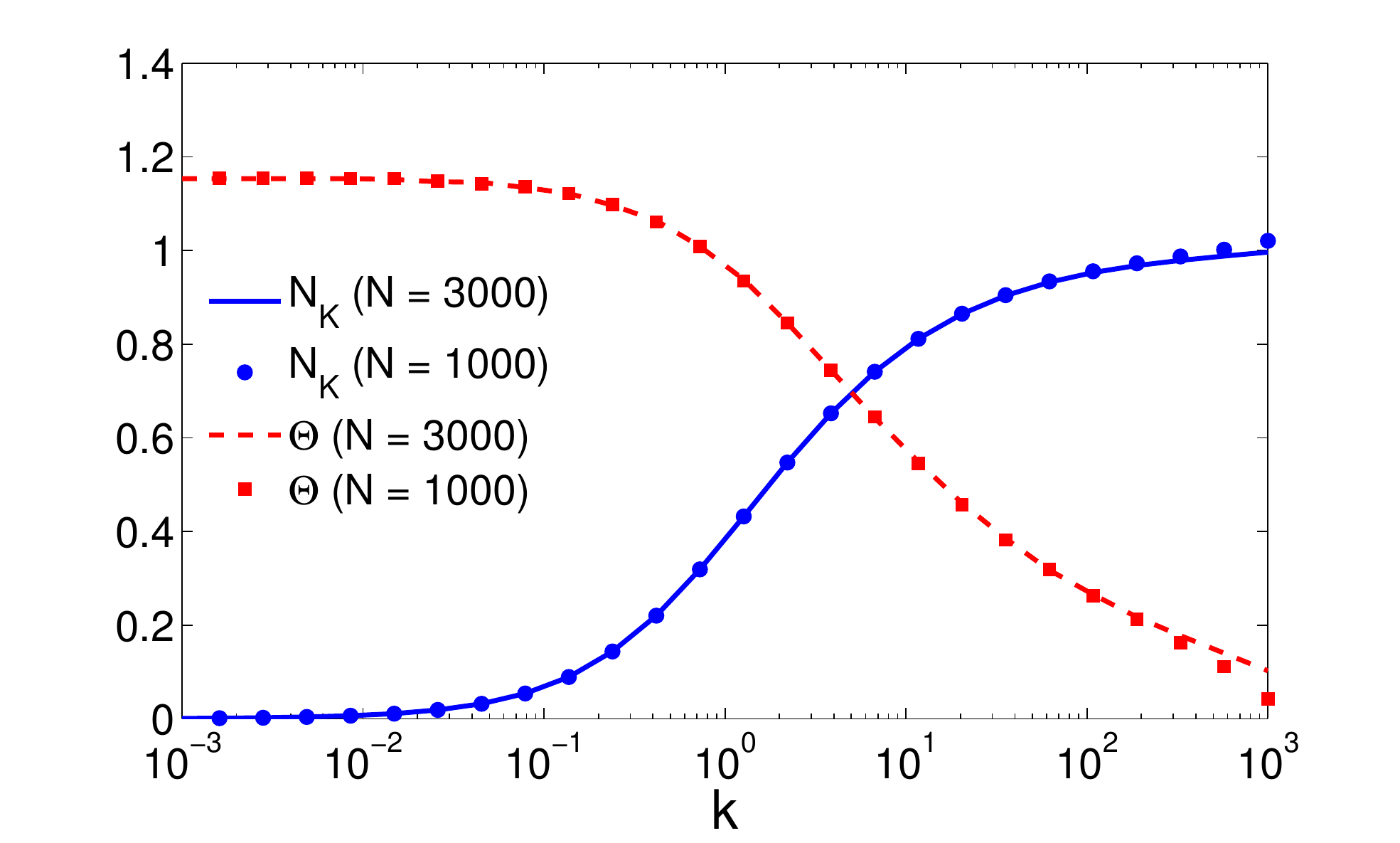}
\caption{
(Color online).  The norm $N_K$ (blue solid line and circles) and the
correlation angle $\Theta$ in radians (red dashed line and squares) as
functions of the adsorption coefficient $k$, for the truncation size
$N = 3000$ (lines) and $N = 1000$ (symbols). }
\label{fig:delta_theta}
\end{figure}

As shown on Fig. \ref{fig:delta_theta}, the spatial correlation is
negligible ($N_K \ll 1, \Theta \approx 1$) as long as $kR < 1$.  In
particular for the reference values $kR = 6.4\cdot 10^{-4}, \ 6.4\cdot
10^{-3}, \ 6.4\cdot 10^{-2}$ used in \cite{berezhkovskii:054115}, the
norm $N_K$ is smaller than $0.05$ and the correlation length is nearly
constant at $\Theta \approx 1.2$.

\begin{figure}[h!]
\centering
\includegraphics[width=90mm]{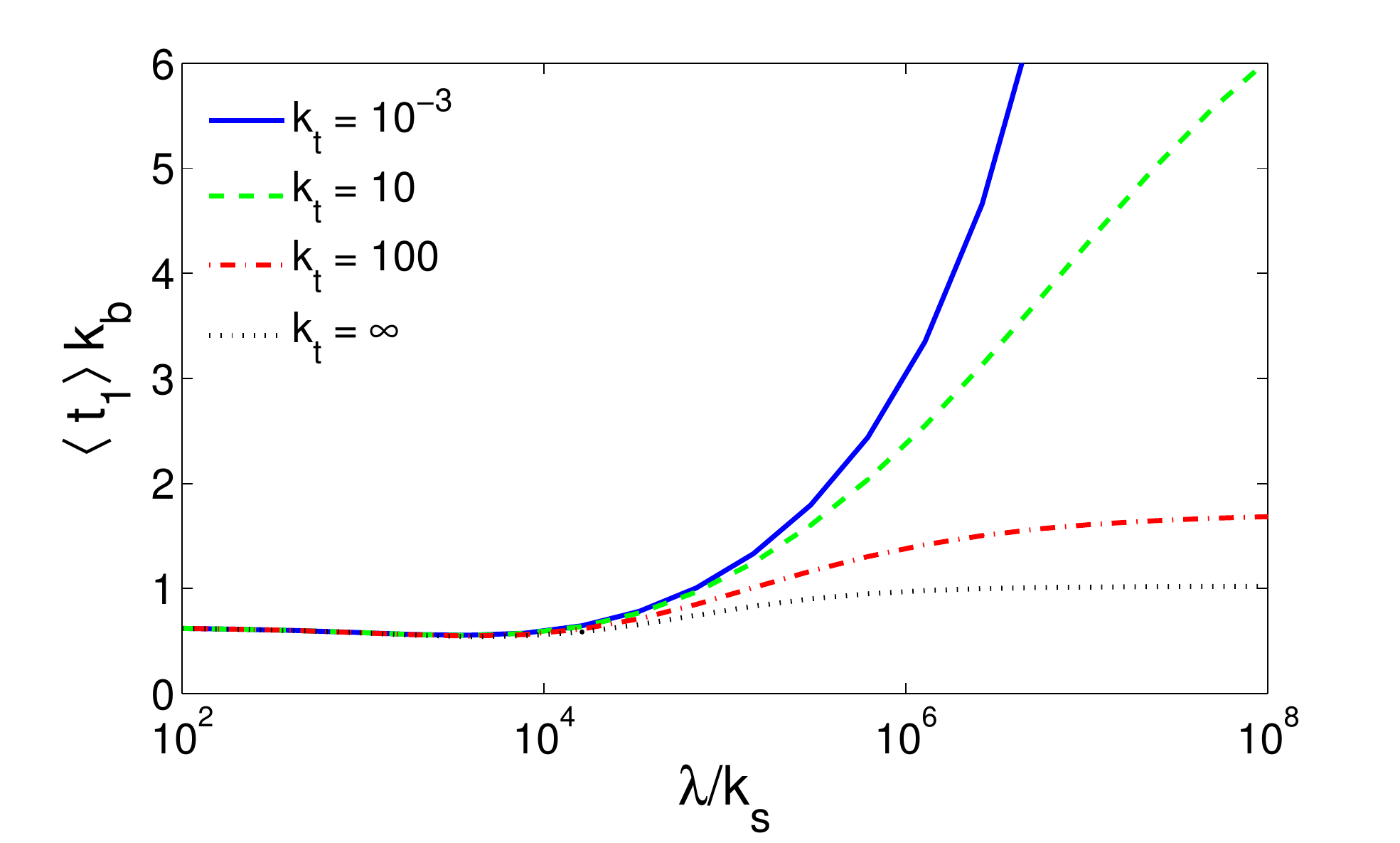}
\caption{
(Color online).  The surface GMFPT $\langle t_1 \rangle$ as a function of the
desorption rate $\lambda$, for several values of the target adsorption
parameter $k_t$ (for particles hitting the target from the bulk), with
$\epsilon = 0.02$ and $D_{2} = D_{1} =1$ in units in which $R=1$.  The
non-reactive part of the surface is semi-reflecting with an adsorption
parameter $k = 100$.  Note that the target is perfectly reactive for
particles adsorbed on the surface.  Series are truncated at $N = 3
\cdot 10^{4}$.}
\label{fig:semireflective}
\end{figure}

\section{Generalization to semi-reflecting targets} \label{sec:semi}

In this section, we briefly generalize our method to the case of a
semi-reflecting target, for which the B.C. reads as
\begin{eqnarray} \label{eq:bcreflectivekt}
t_2(R,\theta) = \begin{cases} \displaystyle - \frac{1}{k_{t}} \pd{t_2}{r}_{\lvert {\bf{r}}=(R,\theta)}  \hskip 14mm (0 \leq \theta \leq \epsilon), \cr
\displaystyle  t_1(\theta) - \frac{1}{k} \pd{t_2}{r}_{\lvert {\bf{r}}=(R,\theta)} \qquad (\epsilon < \theta \leq \pi). \end{cases}
\end{eqnarray}
This general description includes the following cases:
\begin{itemize}
\item 
$k_{t} = \infty$ is the case considered so far in the present article
of a fully adsorbing target, Eq. (\ref{eq:bcreflective});
\item 
$k_{t} = k$ is considered in Ref. \cite{Rupprecht:2012a};
\item 
$k_{t} = 0$ corresponds to a target which is fully reflecting for
particles hitting the target from the bulk.
\end{itemize}

Using the mixed boundary condition (\ref{eq:bcreflectivekt}), the
projection of a series representation (\ref{eq:t2_def_particular}) for
$t_2(R,\theta)$ onto the basis $\{ V_n(\theta)\}_{n\geq 0}$ becomes
\begin{eqnarray} 
&& \int^{\pi}_{0} \left( t_2(R,\theta) + \frac{1}{k_{t}} \pd{t_2}{r}_{\lvert {\bf{r}}
= (R,\theta)}\right) V_{n}(\theta) d\mu(\theta) = \\  \nonumber
&& \int^{\pi}_{\epsilon} t_1(\theta) V_{n}(\theta) d\mu(\theta) -
\int^{\pi}_{\epsilon} \left(\frac{1}{k} - \frac{1}{k_{t}} \right)
\pd{t_2}{r}_{\lvert R} V_{n}(\theta) d\mu(\theta),
\end{eqnarray}
which replaces Eq. (\ref{eq:inteq}), with $d\mu(\theta) = \frac12
\sin\theta d\theta$.  This leads to the following equations on
coefficients $\alpha_{n}$:
\begin{widetext}
\begin{eqnarray}
 \alpha_{0} &=&  \braket{t_1}{1} - \frac{R^{2}}{2 d D_2} \left(1 + 2 \left(\frac{1}{k R} - \frac{1}{k_{t} R}\right) 
K^{(\epsilon)}_{00}\right) -  \sum^{\infty}_{m=1} m \left(\frac{1}{k R} - \frac{1}{k_{t} R}\right) 
K^{(\epsilon)}_{0m} \ \alpha_{m},     \label{eq:sysalpha0_particular_new}\\ 
\alpha_{n} \left(1 + \frac{n}{k_{t} R}\right)  &=& \braket{t_1}{V_{n}} + \frac{R^2}{d D_2} 
\left(\frac{1}{k R} - \frac{1}{k_{t} R}\right) K^{(\epsilon)}_{n0} - 
\sum^{\infty}_{m=1} m \left(\frac{1}{k R} - \frac{1}{k_{t} R}\right) K^{(\epsilon)}_{nm} \ \alpha_{m} \quad (n \geq 1). \label{eq:sysalphan_particular_new}
\end{eqnarray}
\end{widetext}
From these equations, we extend the definition of $M_{nm}$ and
$\hat{U}_{m}$ from Eq. (\ref{def:QandU}) to
\begin{eqnarray} 
 \label{eq:modifiedQ}
 M_{nm}  &\equiv& \delta_{mn} \frac{n}{k_t R} + m \left( \frac{1}{k R} - \frac{1}{k_t R} \right) K^{(\epsilon)}_{nm}, \\
 \label{eq:modifiedU}
 \hat{U}_{n} &\equiv& \braket{t_1}{V_{n}} + \frac{R^2}{d D_2} \left( \frac{1}{k R} - \frac{1}{k_t R} \right) K^{(\epsilon)}_{n0}.
\end{eqnarray}
Following the same steps as in Sec. \ref{sec:integral_eq}, one gets an
integral equation on the dimensionless MFPT $\psi(\theta)$:
\begin{eqnarray} \label{eq:t1distribution_dimless2}
\psi(\theta) &=& g_{\epsilon}(\theta) 
+\sum\limits^{\infty}_{n,m=1} \frac{V_{n}(\theta)-V_{n}(\epsilon)}{\rho_n} ~X_{nm} \\  \nonumber
&& \times \biggl(\frac{R^2}{d D_2 T} \left( \frac{1}{k R} - \frac{1}{k_t R} \right) K^{(\epsilon)}_{0m} + \omega^2 \braket{\psi}{V_{m}} \biggr)  ,
\end{eqnarray}
which generalizes Eq. (\ref{eq:t1distribution_dimless}).  Expanding
this function onto the basis $\{V_n(\theta) - V_n(\epsilon)\}$ yields
Eq. (\ref{eq:dn}), with
\begin{equation}
\label{eq:defU2}
U_{n} \equiv   \frac{1}{\rho_n} \sum^{\infty}_{m=1} X_{nm} \left( \frac{\xi_{m}}{\rho_m} 
+  \frac{R^2}{d D_2} \left( \frac{1}{k R} - \frac{1}{k_t R} \right) \frac{K^{(\epsilon)}_{m0}}{\omega^{2} T}  \right) ,
\end{equation}
which generalizes Eq. (\ref{eq:defU}).  This relation can also be
written as
\begin{equation}
U_n = Z_n + \frac{D_1}{d D_2 (1 + \lambda/\alpha)} \left(\frac{1}{kR} - \frac{1}{k_t R}\right) W_n ,
\end{equation}
where $Z_n$ and $W_n$ are still defined through
Eqs. (\ref{eq:V}, \ref{eq:W}).  Other quantities and representations
remain unchanged.  Repeating the computation of the derivative of
$\langle t_1\rangle$ at $\lambda = 0$, one gets the lower bound as
\begin{equation} \label{eq:lowerbound2}
\left(\frac{D_2}{D_1}\right)_{{\rm low}} = - \frac{1}{d k R} \ \frac{\langle
g_{\epsilon}\rangle + (\bm{\xi} \cdotp \bm{W})(1 - k/k_t)}{(\bm{\xi} \cdotp \bm{Z})} .
\end{equation}
which extends Eq. (\ref{eq:lowerbound}).

One can see that the change in the boundary condition, i.e., extension
from Eq. (\ref{eq:bcreflective}) to Eq. (\ref{eq:bcreflectivekt}),
does not affect the method and the structure of the solution.  As a
consequence, the conclusions on the optimility of the surface GMFPT remain
qualitatively unchanged, although values of the lower bound may be
different.  Note that the determination of the upper bound requires
the expression of the surface GMFPT for a semi-reflecting target in an
otherwise reflecting sphere, which is still unknown.

As shown on Fig. \ref{fig:semireflective}, optimization in $\lambda$
remains possible even in the case of a target which is fully
reflecting for particles hitting the target from the bulk.

\section{Application to a disk}  \label{sec:disk}

\begin{table}[h!]
 \begin{center}
 \begin{tabular}{|c|c|}
  \hline
$V_{n}(\theta)$ & $ \sqrt{2} \cos(n \theta)$\\
  \hline
$g_{\epsilon}(\theta)$ & $\frac{1}{2} (\theta-\epsilon)(2\pi-\epsilon-\theta)$  \\
  \hline
$\langle g_{\epsilon}|1 \rangle_{\epsilon} \equiv \langle g_{\epsilon} \rangle_{\epsilon} $ 
& $\frac{1}{3 \pi} (\pi-\epsilon)^3$ \\
  \hline
$\xi_{n}~ (n \geq 1)$ & 
$- \frac{\sqrt{2}}{ \pi} \{ (\pi - \epsilon) \cos(n \epsilon) + \sin(n \epsilon)/n \}$ \\
  \hline
$K^{(\epsilon)}_{00}$ & $ (\pi - \epsilon)/\pi $\\
  \hline
$K^{(\epsilon)}_{n0}$ ~ $(n\geq 1)$  & $ -\frac{\sqrt{2}}{n \pi} \sin(n\epsilon) $\\
  \hline
$K^{(\epsilon)}_{nn}$ ~ $(n\geq 1)$  & 
$ \frac{1}{\pi} \left( \pi - \epsilon - \frac{\sin(2 n \epsilon)}{2 n} \right)  $\\
 \hline
$K^{(\epsilon)}_{nm} ~ \quad n,m \geq 1, m \neq n$ & $ \frac{1}{\pi} \left( \frac{\sin((m-n)\epsilon)}{m-n} + \frac{\sin((m+n)\epsilon)}{m+n}\right)$ \\
  \hline
$I^{(\epsilon)}_{nn} ~ (n\geq 1)$ & 
$ \frac{1}{\pi} \left(\pi - \epsilon + \frac{\sin 2 n \epsilon}{2 n} \right)$  \\
  \hline
$I^{(\epsilon)}_{nm} \quad m,n\geq 1, m \neq n$ & 
$\frac{2}{\pi} \frac{\cos(n\epsilon) \frac{\sin (m\epsilon)}{m} - \cos (m\epsilon) 
\frac{\sin (n\epsilon)}{n}}{n^2 - m^2} ~m^2$ \\
\hline
\end{tabular}
 \end{center}
\caption{
Summary for the 2D case of the quantities involved in the computation
of the vector $\bm{\xi}$ and the matrices $Q$ and $M$ in
Eqs. (\ref{eq:defU}, \ref{eq:defQ}) that determine the Fourier
coefficients $d_n$ of $t_1(\theta)$ according to Eq. (\ref{eq:dn}).}
\label{tab:Vtable2D}
\end{table}

In 2D, the diffusion equations on the MFPT are identical to
Eqs. (\ref{eq:temps1}, \ref{eq:temps2}) provided the change of
definition for the Laplace operators:
\begin{equation*}
\Delta_{r} = \pdds{}{r} + \frac{1}{r} \pd{}{r} , \qquad  \Delta_{\theta} = \pa^{2}_{\theta}.
\end{equation*}
The eigenfunctions $V_{n}(\theta)$ and eigenvalues $\rho_n$ of the
angular Laplace operator $\Delta_{\theta}$ become
\begin{equation}
V_n(\theta) = \sqrt{2} \cos n\theta, \qquad \rho_n = n^2 \qquad (n\geq 0) .
\end{equation}
The inner scalar product (\ref{eq:scalrproduct}) is replaced by
\begin{equation} 
(f,g) \rightarrow \langle f | g \rangle_{\epsilon} \equiv \frac{1}{\pi} \int^{\pi}_{\epsilon} f(\theta) g(\theta) d\theta.
\end{equation}
In Table \ref{tab:Vtable2D}, we summarize the expressions for the
matrix elements $K^{(\epsilon)}_{mn}$ and $I^{(\epsilon)}_{mn}$ which
replace those from Table \ref{tab:Vtable}.  One has also to replace
Eq. (\ref{eq:g_eps}) by
\begin{equation}
g_\epsilon(\theta) = \frac12 (\theta - \epsilon)(2\pi - \epsilon - \theta) .
\end{equation}

\begin{figure}[h!]
 \centering
\includegraphics[width=90mm]{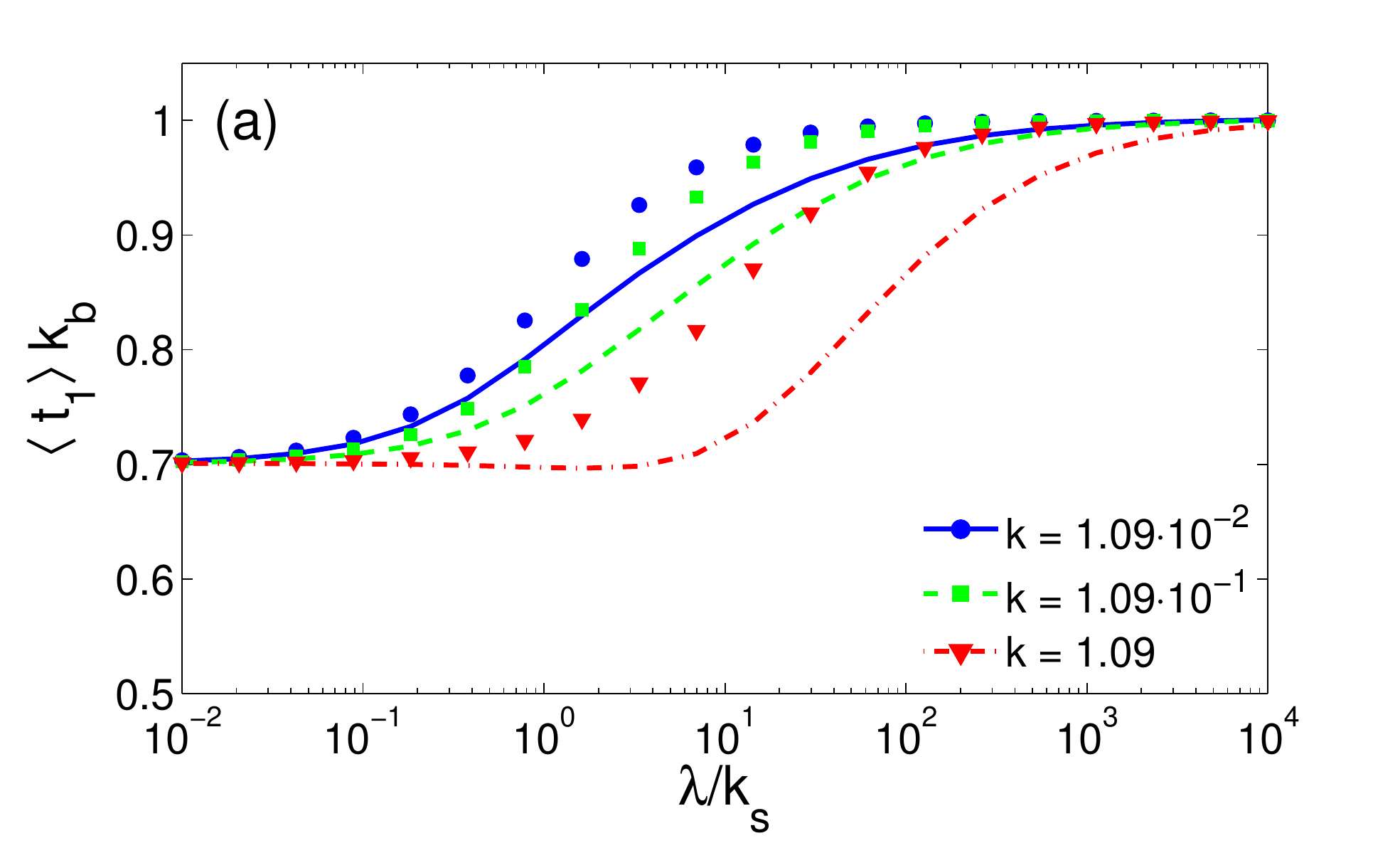}
\includegraphics[width=90mm]{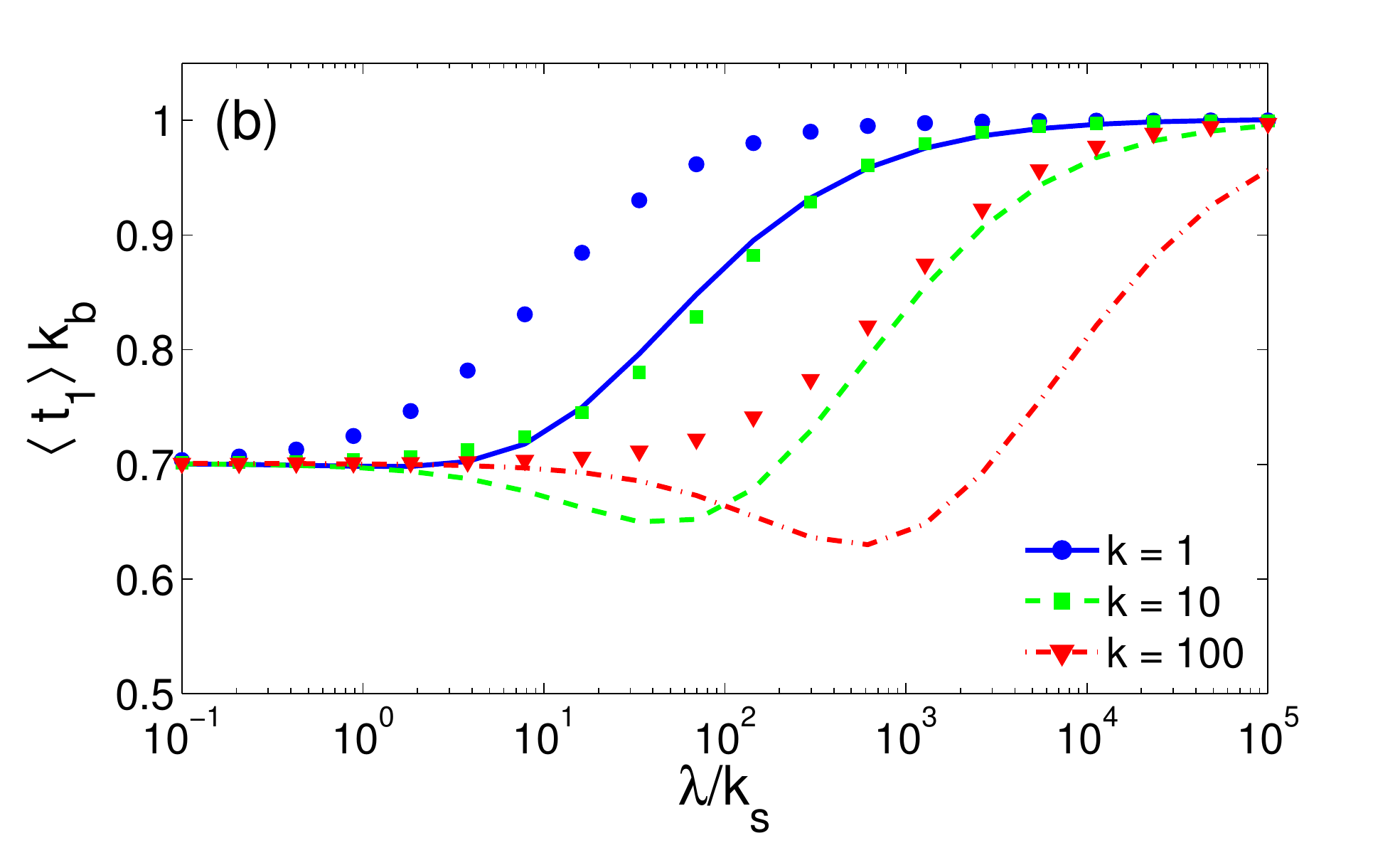}
\caption{
(Color online).  The surface GMFPT $\langle t_1 \rangle$ in the disk as a
function of the desorption rate $\lambda$, for $\epsilon = 0.02$,
$D_2/D_1 = 1$, and {\bf (a)} $kR = 1.09 \cdot 10^{-2},\ 1.09 \cdot
10^{-1},\ 1.09$ (corresponding to $\alpha = 0.1 k_b, k_b, 10 k_b$) and
{\bf (b)} $kR = 1,\ 10,\ 100$.  Series are truncated at $N = 3 \cdot
10^{4}$. }
\label{fig:mfpt2D}
\end{figure}

\begin{figure}[h!]
\centering
\includegraphics[width=90mm]{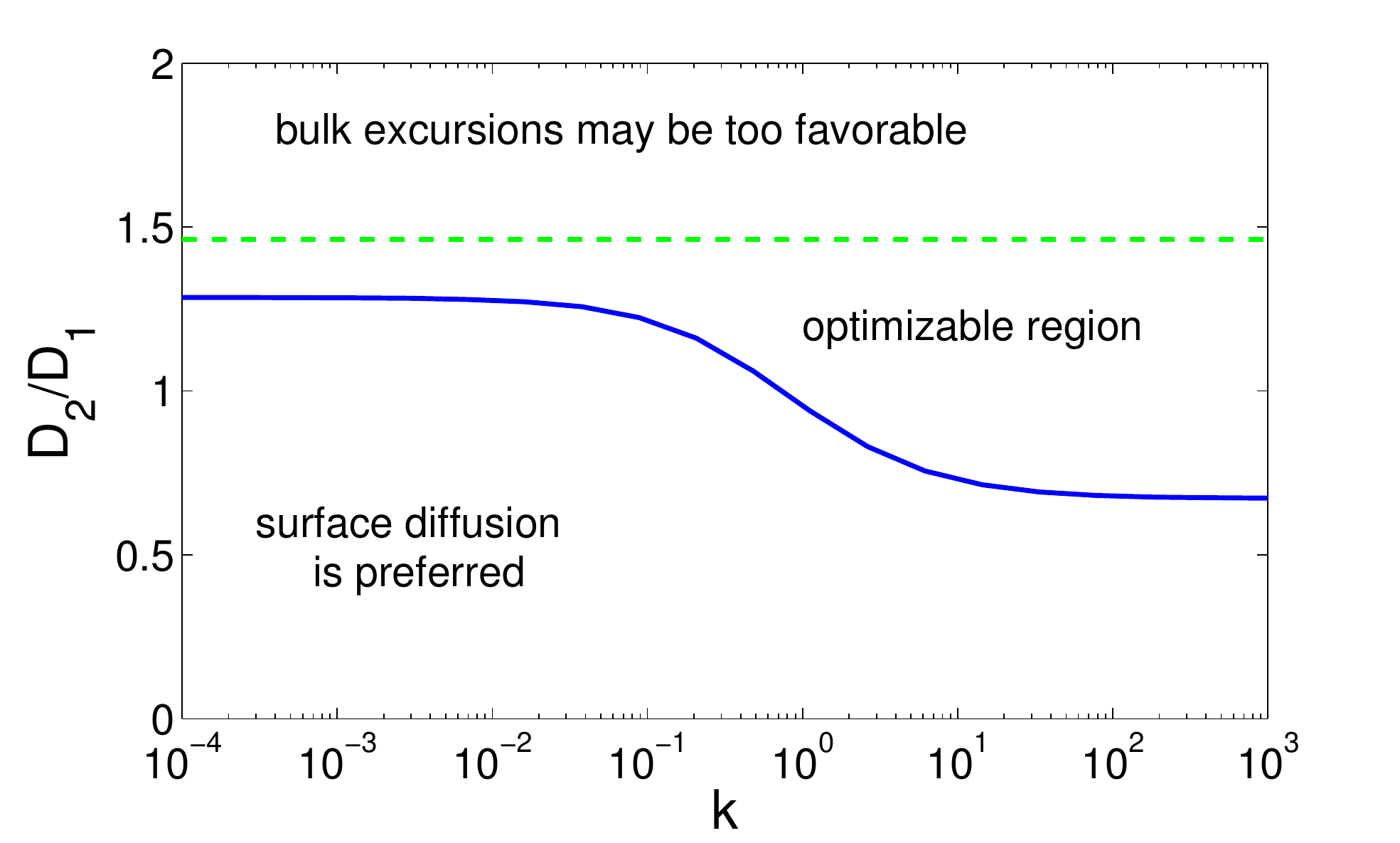}
\caption{
(Color online).  The regions of optimality for the surface GMFPT $\langle
t_1\rangle$ for a disk of radius $R = 1$ with an aperture of
half-width $\epsilon = 0.02$.  Below the lower bound (solid red line)
surface diffusion is preferred.  Above the upper bound (dashed green
line), the surface GMFPT is higher in the adsorbed state than in the desorbed
state.  In between, the surface GMFPT is an optimizable function of $\lambda$.
Series are truncated at $N = 10^{4}$.}
\label{fig:D2vsk_2D}
\end{figure}

The method of derivation and the remaining quantities are not
modified.  In particular, Eq. (\ref{eq:t1distribution}) for the MFPT
$t_1(\theta)$, Eq. (\ref{eq:searchtime}) for the surface GMFPT $\langle
t_1\rangle$ and Eq. (\ref{eq:lowerbound}) for the lower bound on the
diffusion coefficient ratio $\left(D_2/D_1\right)_{{\rm low}}$ are
applicable for the disk.  In turn, the asymptotic relations on the
exit time (\ref{eq:taus3D}, \ref{eq:taub3D}) are modified:
\begin{eqnarray}
t_s &=& \langle t_1\rangle_{\lambda=0} = \frac{R^2}{D_1} ~ \frac{(\pi-\epsilon)^3}{3\pi} , \\
t_b &=& \langle t_1\rangle_{\lambda\to\infty} \simeq \frac{R^2}{D_2} ~ \biggl[\ln (2/\epsilon) + O(1)\biggr]
\end{eqnarray}
that leads to the following expression for the upper bound on the
diffusion coefficients
\begin{equation}
  \left(\frac{D_2}{D_1}\right)_{{\rm up}} = \frac{3\pi \ln(2/\epsilon)}{(\pi-\epsilon)^3} + O(1).
\end{equation}

Figure \ref{fig:mfpt2D} illustrates the fact that the surface GMFPT is an
optimizable function of the desorption rate $\lambda$ in the range of
parameters represented on Fig. \ref{fig:D2vsk_2D}.

\end{document}